\documentclass[twocolumn]{aastex63}

\usepackage{aas_macros}
\usepackage[lined]{algorithm2e}
\usepackage{lineno}
\usepackage{verbatim, amsmath, amsfonts, amssymb,amssymb}
\usepackage{pifont}
\usepackage{enumitem}
\usepackage{booktabs}
\usepackage{multirow}
\usepackage[utf8]{inputenc}
\usepackage[T1]{fontenc}
\usepackage{empheq}
\usepackage[skins,theorems]{tcolorbox}
\usepackage{latexsym}
\usepackage{bigints}
\usepackage{natbib}
\usepackage{subfigure, apalike}
\usepackage{color}
\usepackage{tikz}
\usepackage{lineno}
\tcbset{highlight math style={enhanced,
 colframe=cyan!10!gray,colback=gray!5!white,arc=4pt,boxrule=1pt,
  drop fuzzy shadow}}

\newenvironment{alignteo}%
  {\empheq[box=\tcbhighmath]{align}}
  {\endempheq}
  
\newcommand{\e}[1]{\ensuremath{\times 10^{#1}}}
\newcommand{\cmark}{\ding{51}}%
\newcommand{\xmark}{\textcolor{lightgray}{\ding{55}}}

\newcommand{\ddn}{{D(d,n)$^3$He}}

\newcommand{\dpg}{{D(p,$\gamma$)$^3$He}}

\definecolor{red2}{HTML}{7f0000}

\def\iso#1#2{\mbox{${}^{#2}{\rm #1}$}}
\def\h#1{\iso{H}{#1}}

\received{\today}
\submitjournal{ApJ}

\begin{document}

\title{Bayesian Estimation of the D(p,$\gamma$)$^3$He Thermonuclear Reaction Rate}

\author{Joseph Moscoso}
\affiliation{Department of Physics \& Astronomy, University of North Carolina at Chapel Hill, NC 27599-3255, USA}
\affiliation{Triangle Universities Nuclear Laboratory (TUNL), Durham, North Carolina 27708, USA}

\author[0000-0001-7207-4584]{Rafael S. de Souza}
\affiliation{Key Laboratory for Research in Galaxies and Cosmology, Shanghai Astronomical Observatory,\\ Chinese Academy of Sciences, 80 Nandan Road, Shanghai 200030, China}

\author{Alain Coc}
\affiliation{CNRS/IN2P3, IJCLab, Universit\'e Paris-Saclay, B\^atiment, 104, F-91405 Orsay Campus, France}

\author[0000-0003-2381-0412]{Christian Iliadis}
\affiliation{Department of Physics \& Astronomy, University of North Carolina at Chapel Hill, NC 27599-3255, USA}
\affiliation{Triangle Universities Nuclear Laboratory (TUNL), Durham, North Carolina 27708, USA}

\color{red}
\begin{abstract}
Big bang nucleosynthesis (BBN) is the standard model theory for the production of the light nuclides during the early stages of the universe, taking place for a period of about 20 minutes after the big bang. Deuterium production, in particular, is highly sensitive to the primordial baryon density and the number of neutrino species, and its abundance serves as a sensitive test for the conditions in the early universe. The comparison of observed deuterium abundances with predicted ones requires reliable knowledge of the relevant thermonuclear reaction rates, and their corresponding uncertainties. Recent observations reported the primordial deuterium abundance with percent accuracy, but some theoretical predictions based on BBN are at tension with the measured values because of uncertainties in the cross section of the deuterium-burning reactions. In this work, we analyze the S-factor of the D(p,$\gamma$)$^3$He reaction using a hierarchical Bayesian model. We take into account the results of eleven experiments, spanning the period of 1955--2021; more than any other study. We also present results for two different fitting functions, a two-parameter function based on microscopic nuclear theory and a four-parameter polynomial. Our recommended reaction rates have a 2.2\% uncertainty at $0.8$~GK, which is the temperature most important for deuterium BBN. Differences between our rates and previous results are discussed.
\end{abstract}
\color{black}




\section{Introduction} 
\label{sec:intro}
Apart from the universal expansion and the cosmic microwave background (CMB) radiation, the third observational evidence for the hot big bang model derives from big bang (i.e., primordial) nucleosynthesis (BBN). The detailed comparison between the measured and predicted primordial abundances of the light elements ($^4$He, D, $^3$He, and $^{7}$Li) provides a sensitive test of the BBN model because the abundances span a range of nine orders of magnitude. This theory is parameter-free, since we know from the $Z^0$ lineshape measured by CERN's Large Electron-Positron (LEP) collider that the number of light neutrino families is three. We also know, with 0.62\% precision, the baryonic density of the universe, $\Omega_{\mathrm{b}}{\cdot}h^2 = 0.02242 \pm 0.00014$, from the measurement of the CMB anisotropies by the \cite{collaboration2018planck}, combined with baryon acoustic oscillations (BAO) data \citep{BAO17}.

Over the past years, measured primordial D and $^4$He abundance values have been reported with precision reaching percent levels. The deuterium primordial abundance is determined from observations of cosmological clouds at high redshift on the line of sight of distant quasars. Recently, new observations and the reanalysis of existing data by \cite{Cooke2018} resulted in an average value of D/H $=$ $(2.527\pm 0.030)\times10^{-5}$ (with 1.2\% precision). The primordial abundance of $^4$He is deduced from observations in H{\sc ii} (ionized hydrogen) regions of compact blue galaxies. \cite{Aver2015} obtained a $^4$He nucleon fraction of $Y_p$ $=$ $0.2449\pm 0.0040$, which was recently updated to a value of $Y_p$ $=$ $0.2453\pm0.0034$ (with 1.4\% precision) by including data from the extremely metal-poor galaxy Leo P \citep{Ave21}. The primordial abundances of $^3$He and $^7$Li do not provide useful constraints for BBN at this time. The nuclide $^3$He is both produced and destroyed by stars and, therefore, its abundance evolution is not well understood. The measured $^7$Li abundance is a factor of $\approx3$ smaller compared to the BBN predictions \citep[e.g.,][]{Fie11,Pitrou2018}. Although this cosmological lithium problem has not found a satisfactory solution yet, the present consensus is that it is most likely not caused by erroneous nuclear physics input \citep{Dav20,IC20}.
 
In this work, we will focus on the more recently discovered {\it tension} between the measured and predicted primordial D/H values \citep{Coc15,Pitrou2018,IC20}. Less than a dozen nuclear reactions are important for the BBN of the light elements, and only five are important for deuterium production. Deuterium is directly produced through the p(n,$\gamma$)D reaction and destroyed by the D(p,$\gamma)^3$He, D(d,n)$^3$He, and D(d,p)$^3$H reactions. The weak interactions, n$\leftrightarrow$p, that convert neutrons to protons, and vice versa, can be precisely calculated by using as experimental input either the neutron lifetime $\tau_{\rm n} = 879.4 \pm 0.6\,{\rm s}$ \citep{PDG2020} or, alternatively, the quark mixing angle $V_{\rm ud}$ and the nucleon axial coupling constant $g_A$ \citep{Pitrou2018,Pit21}. The uncertainty in the weak rates is $\approx$0.06\%, which will impact the $^4$He abundance, but has no significant effect on the deuterium abundance. The p(n,$\gamma$)D reaction rate is obtained from effective field theory \citep{Ando2006} with an uncertainty of $<$1\% at BBN temperatures. Experimental D(d,n)$^3$He and D(d,p)$^3$H reaction rates have recently been reanalyzed using hierarchical Bayesian models \citep{gomez17}, resulting in uncertainties of about $\approx$1\%. The remaining reaction rate, D(p,$\gamma)^3$He, has also been reanalyzed using a Bayesian model \citep{iliadis16}. It was found to have a much larger uncertainty (3.7\% at 0.8~GK) compared to the other processes involving deuterium, which impacts the predicted primordial deuterium abundance significantly. 

\begin{figure}
\includegraphics[width=0.975\linewidth]{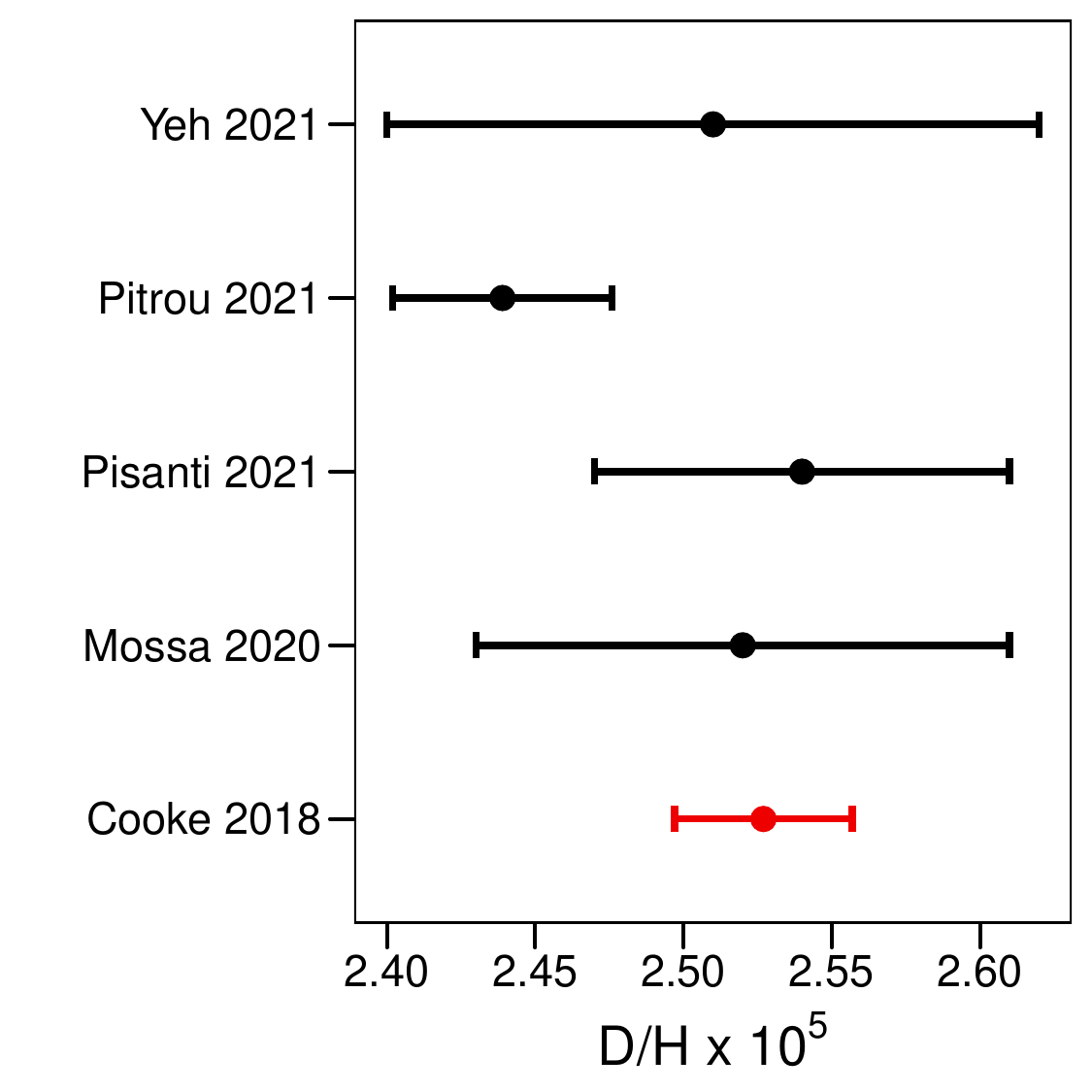}
\caption{Primordial D/H number abundance ratio from BBN predictions (black) and observations (red). References: \citet{Cooke2018} ($2.527\pm 0.030$); \citet{Mossa20} ($2.52\pm0.03\pm0.06$); \citet{Pit21} ($2.439\pm0.037$);  \citet{Pis21} ($2.54\pm0.07$); \citet{Yeh21} ($2.51\pm0.11$). Uncertainties refer to combined standard uncertainties (i.e., 1$\sigma$).}
\label{fig:DoverH}
\end{figure}

Recently, new D(p,$\gamma)^3$He cross section data for the energy range important for BBN ($20$ $-$ $300$~keV in the d+p center of mass) have been reported by three experiments \citep{Tisma:2019ug,Mossa20,Turkat21}. Based on the first two works, updated rates have been adopted in simulations to study the BBN production of deuterium \citep{Mossa20,Pit21,Pis21,Yeh21}. Results are displayed in Figure~\ref{fig:DoverH}. It is apparent that the predicted D/H values of \citet{Mossa20}, \citet{Pis21}, and \citet{Yeh21} agree with the observed value \citep{Cooke2018}, although the uncertainties on the predicted abundances significantly exceed those on the measured abundance. On the other hand, the more precise prediction of \citet{Pit21} (see also \citet{Coc15} and Figure~3 in \citet{IC20}) displays an $\approx1.8\sigma$-tension with the observations. If confirmed, it could have important implications pointing to new physics beyond the standard model.

The BBN abundance predictions from modern public codes, such as {\tt PArthENoPE} \citep{Parthenope2,Parthenope3} or {\tt PRIMAT} \citep{Pitrou2018}, or private codes as in \citet{Yeh21},
essentially differ only by the set of adopted thermonuclear reaction rates.
Since a precision on the order of 1\% is required for all reaction rates impacting the light nuclide abundances, the goal of the present work is to reanalyze the important D(p,$\gamma)^3$He rates by taking into account the latest experimental results \citep{Tisma:2019ug,Mossa20,Turkat21} and adopting a hierarchical Bayesian model.

In Section~\ref{sec:strategies} we will discuss data selection, statistical models, and fitting functions. Section~\ref{sec:Bayemodel} introduces our Bayesian model. Results are discussed in Section~\ref{sec:results}. Section~\ref{sec:rates} presents our new reaction rates, together with a comparison to previous results. A concluding summary is provided in Section~\ref{sec:summary}. In the Appendix, we discuss the data considered for the analysis.

\section{Strategies} 
\label{sec:strategies}
Before presenting our analysis, we find it necessary to comment on a few issues that are not always spelled out in the recent literature, and which have caused some confusion. 

Nonresonant thermonuclear reaction rates are computed from the astrophysical S-factor (see below). Generally, the S-factor can be estimated by two different methods. First, if no experimental data exist, it can be calculated based purely on some nuclear reaction model (``theory-based S-factor''). Second, if experimental data exist, an experimental S-factor can be estimated by using a statistical model that will, by necessity, employ some fitting function (``experiment-based S-factor''). Reaction rates for D(p,$\gamma$)$^3$He that are obtained from a theory-based S-factor are shown, e.g., as green lines in Figure~3 of \citet{Pis21} or Figure~2 of \citet{Yeh21}. Since experimental data exist for the D(p,$\gamma$)$^3$He reaction, few researchers would adopt a theory-based S-factor in their BBN simulations.

Because the goal for the D(p,$\gamma$)$^3$He reaction is the estimation of an experiment-based S-factor, we need to discuss the assumptions that can lead to different results obtained by different groups using different methods of analysis. Three main sources may cause such differences. First, the data sets to be analyzed need to be selected. Second, a statistical model for the data analysis must be chosen. Third, a function to be fitted in the statistical analysis has to be adopted. We will discuss these issues sequentially.

\subsection{Data selection} 
\label{sec:selection}
Different kinds of nuclear reaction data can be used in the statistical analysis. The best ones to use pertain to cross sections or S-factors obtained in direct measurements of the reaction of interest, i.e., D(p,$\gamma$)$^3$He, over a wide region of energies which include the range of importance for BBN ($20$ $-$ $300$~keV in the d $+$ p center of mass system). Another possibility is to adopt data from the reverse reaction, $^3$He($\gamma$,p)D, and apply the reciprocity theorem to calculate the cross section or S-factor for D(p,$\gamma$)$^3$He. Indirect measurements are a third possibility; i.e., data obtained in a reaction (e.g., proton transfer) other than the one of direct interest, from which the D(p,$\gamma$)$^3$He S-factor can be deduced by applying some nuclear reaction model. For the D(p,$\gamma$)$^3$He reaction, most BBN studies have focused on the first two possibilities. A major problem with the third possibility is the difficulty of reliably estimating the systematic uncertainties involved.

Table~\ref{tab:references} provides an overview of the nuclear data sets (listed in the first column) used in the present and previous D(p,$\gamma$)$^3$He rate estimates (top row). All of the listed experiments represent direct measurements of the D(p,$\gamma$)$^3$He reaction at astrophysically important energies. The only exception is the work of \citet{War63}, which represents the measurement of the reverse reaction, $^3$He($\gamma$,p)D. For the works of \citet{War63,Sch97,Ma97,Cas02,Tisma:2019ug,Mossa20,Turkat21}, sufficient information is provided to estimate both the statistical and systematic uncertainties of the S-factor. These ``absolute data'' represent the most important data sets because their absolute normalizations impact the magnitude of the fitted S-factor. For the experiments of \citet{Gri55,Gri62,Gri63,Bai70}, either no uncertainties or only combined (statistical and systematic) uncertainties are presented in the original works. Although we do not trust the absolute normalization of the latter data sets, they do contain valuable information on the energy dependence of the S-factor. Therefore, we will implement them into our Bayesian model as ``relative data'' only, as will be explained in Section~\ref{sec:Bayemodel}. 

\begin{deluxetable*}{lcccc} 
\label{tab:references}
\tablecaption{D(p,$\gamma$)$^3$He data sets (first column, listed in chronological order) taken into account in various BBN studies (top row).\tablenotemark{a}}
\tablewidth{\linewidth}
\tablehead{
Reference           &   Present     & \citet{Pitrou2018}\tablenotemark{a} &  \citet{Yeh21}  &  \citet{Pis21}\tablenotemark{b} 
         } 
\startdata
   \cite{Turkat21}\tablenotemark{c}     &   \cmark                  &  {\textcolor{gray}{\xmark}}       &   {\textcolor{gray}{\xmark}}  & {\textcolor{gray}{\xmark}} \\
   \cite{Mossa20}\tablenotemark{c}      &   \cmark                  &  {\textcolor{gray}{\xmark}}       &   \cmark                      &  \cmark    \\ 
   \cite{Tisma:2019ug}\tablenotemark{c} &   \cmark                  &  \xmark                           &   \xmark                      &  \cmark    \\ 
   \cite{Bys08}                         &   \xmark                  &  \cmark                           &   \xmark                      &  \xmark   \\ 
   \cite{Cas02}\tablenotemark{c}        &   \cmark                  &  \cmark                           &   \cmark                      &  \cmark    \\ 
   \cite{Sch97}\tablenotemark{c}        &   \cmark                  &  \cmark                           &   \cmark                      &  \cmark    \\ 
   \cite{Ma97}\tablenotemark{c}         &   \cmark                  &  \cmark                           &   \cmark                      &  \cmark    \\ 
   \cite{Bai70}                         &   \cmark\tablenotemark{e} &  \xmark                           &   \xmark                      &  \xmark   \\ 
   \cite{Wol67}                         &   \xmark                  &  \xmark                           &   \cmark                      &  \xmark   \\ 
   \cite{Gel67}                         &   \xmark                  &  \xmark                           &   \xmark                      &  \cmark    \\ 
   \cite{War63}\tablenotemark{c,d}      &   \cmark                  &  \xmark                           &   \xmark                      &  \cmark    \\ 
   \cite{Gri63}                         &   \cmark\tablenotemark{e} &  \xmark                           &   \xmark                      &  \cmark    \\ 
   \cite{Gri62}                         &   \cmark\tablenotemark{e} &  \xmark                           &   \cmark                      &  \cmark    \\ 
   \cite{Gri55}                         &   \cmark\tablenotemark{e} &  \xmark                           &   \xmark                      &  \xmark   \\ 
\enddata
\tablenotetext{a}{The D(p,$\gamma$)$^3$He rate was adopted from \citet{iliadis16}.}
\tablenotetext{b}{Data sets selected for analysis are identical to those of \citet{Mossa20}, except that the latter work disregarded the experiment of \cite{Tisma:2019ug}.}
\tablenotetext{c}{Statistical and systematic uncertainties of the experimental S-factor have been reported separately.}
\tablenotetext{d}{Measurement of reverse reaction, $^3$He($\gamma$,p)D.}
\tablenotetext{e}{Results implemented into our Bayesian model by treating them as relative S-factors only (see main text).}
\end{deluxetable*}

All of these data just discussed apply to center-of-mass energies below $2$~MeV. Since deuterium synthesis occurs at BBN energies between $20$~keV and $300$~keV, data at energies above $2$~MeV are irrelevant for the considerations of the present work. We disregarded three of the experiments listed in Table~\ref{tab:references} \citep{Wol67,Gel67,Bys08} because they provide too little information to extract reliable S-factors. All of these experiments are discussed in more detail in Appendix~\ref{sec:app}. The appendix also lists, for certain data sets, numerical S-factor values if they had not been reported previously.

\begin{figure*}[hbt!]
\centering
\includegraphics[width=0.95\linewidth]{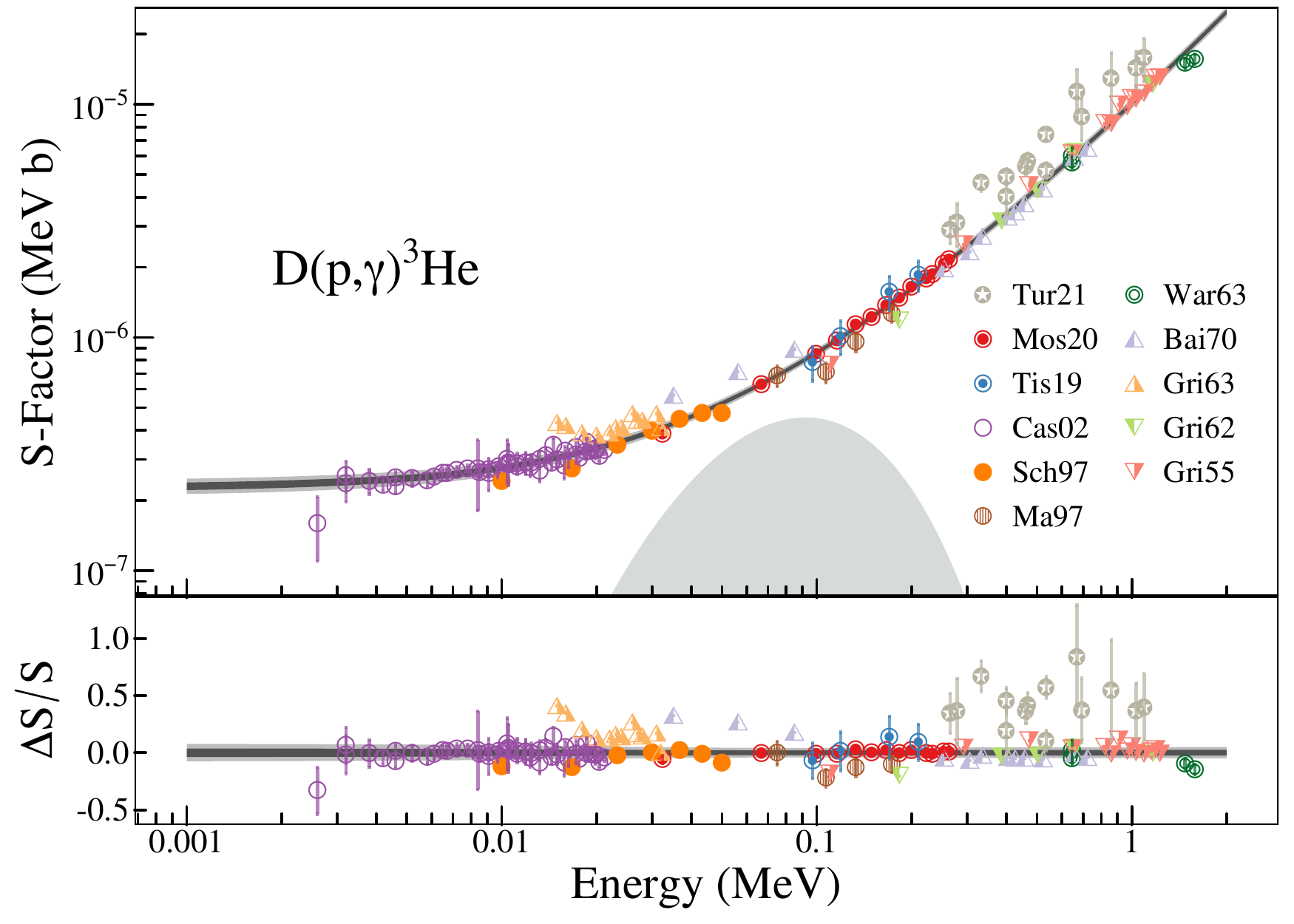}
\caption{(Top) S-factor fit and data for D(p,$\gamma$)$^3$He. The two-parameter fit function was $S(E)$ $=$ (Scale factor) $\times$ $S_{theory}$ $+$ (Offset) MeVb, where $S_{theory}$ denotes the theoretical S-factor of \citet{marcucci05}. The dark and light gray shaded bands correspond to 68 and 95 percentiles, respectively, of the predicted S-factor. Circles depict data for which we adopt absolute normalizations, while triangles refer to relative data. Displayed error bars refer to statistical uncertainties for the absolute data only. The gray-shaded region shows the Gamow peak at a temperature of $T$ $=$ $0.8$~GK. (Bottom) Residuals of fit or data with respect to the median (50 percentile) S-factor at each energy, i.e., $\Delta S/S$ $\equiv$ $(S_{fit~or~data}$ $-$ $S_{median})/S_{median}$. Data set key: (Tur21) \citet{Turkat21}; (Mos20) \citet{Mossa20}; (Tis19) \citet{Tisma:2019ug}; (Cas02) \citet{Cas02}; (Sch97) \citet{Sch97}; (Ma97) \citet{Ma97}; (War63) \citet{War63}; (Bai70) \citet{Bai70}; (Gri63) \citet{Gri63};  (Gri62) \citet{Gri62}; (Gri55) \citet{Gri55}.
}
\label{fig:sfactor}
\end{figure*}

\begin{figure*}[hbt!]
\centering
\includegraphics[width=0.95\linewidth]{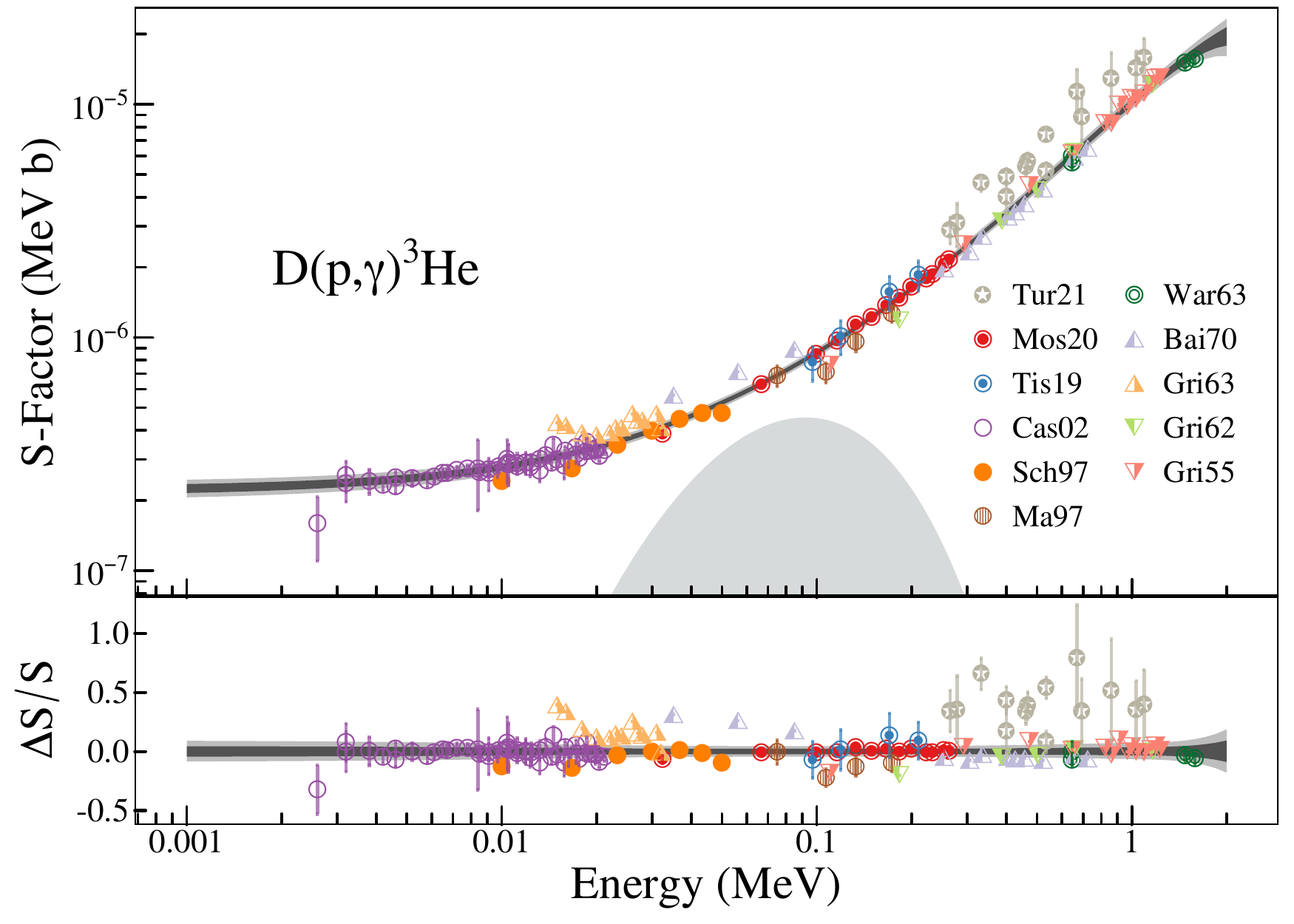}
\caption{(Top) S-factor fit and data for D(p,$\gamma$)$^3$He. The four-parameter fit function was $S(E)$ $=$ $S_0$ $+$ $S_1 E$ $+$ $S_2 E^2$ $+$ $S_3 E^3$ MeVb. The dark and light gray shaded bands correspond to 68 and 95 percentiles, respectively, of the predicted S-factor. Circles depict data for which we adopt absolute normalizations, while triangles refer to relative data. Displayed error bars refer to statistical uncertainties for the absolute data only. The gray-shaded region shows the Gamow peak at a temperature of $T$ $=$ $0.8$~GK. (Bottom) Residuals of fit or data with respect to the median (50 percentile) S-factor at each energy, i.e., $\Delta S/S$ $\equiv$ $(S_{fit~or~data}$ $-$ $S_{median})/S_{median}$. For the data set key, see caption of Figure~\ref{fig:sfactor}. 
}
\label{fig:sfactor2}
\end{figure*}

In total, we take eleven experiments into account in our Bayesian analysis, more than any other BBN study. These data sets are displayed in Figures~\ref{fig:sfactor} and \ref{fig:sfactor2}. The circles depict data for which we adopt absolute normalizations, while triangles refer to relative data. Displayed error bars refer to statistical uncertainties for the absolute data only. In some cases, the error bar is smaller than the symbol size.

The D(p,$\gamma$)$^3$He rate used by \citet{Pitrou2018} was adopted from \citet{iliadis16} and rests on just four experiments \citep{Sch97,Ma97,Cas02,Bys08}. Their strategy was to include only those data sets that reported both statistical and systematic uncertainties, while the two most recent measurements \citep{Tisma:2019ug,Mossa20} had not been published yet. Notice that, in the present study, we disregarded the data of \cite{Bys08} which were taken into account in the previous evaluations of \citet{Coc15,iliadis16}. Our reasoning is explained in Appendix~\ref{sec:ref_not}. 

The BBN study of \citet{Yeh21} considered six experiments and disregarded the works of \citet{Gri55,War63,Gri63,Gel67,Bai70,Bys08,Tisma:2019ug}. They neither used the results of \citet{Bai70} nor \citet{Gri63} {\it ``due to systematic issues with the stopping powers.''} We concur with their explanation (see Appendix~\ref{sec:app}), but instead we include these data as relative S-factors, as will be explained below. 

The BBN study of \citet{Pis21} took nine experiments into account, disregarding the works of \citet{Gri55,Wol67,Bai70,Bys08}. It can be seen from Table~\ref{tab:references} that their data selection is nearly identical to that of the present work, although their analysis differs from ours (see below). 

The very recently published S-factors of \cite{Turkat21} could not be considered in any of the above BBN studies.

\subsection{Statistical models} 
\label{sec:statmodel}
Since all data are subject to known and unknown sources of uncertainties, it
is of utmost importance for a rigorous analysis to devise a method taking these effects into account, while introducing the least amount of bias. Prior to 2016, the statistical models used for analyzing D(p,$\gamma$)$^3$He data were exclusively based on some variant of the $\chi^2$ minimization procedure outlined in \citet{dagostini}. Such models are also used in the recent works of \citet{Yeh21,Pis21}. {\bf Chi-square minimizations make the implicit assumption that all sources of uncertainties entering the model can be described by Gaussian distributions, but this purely Gaussian assumption frequently does not apply in practice.}


{\bf For the case of thermonuclear reaction rates in general}, one needs to consider statistical and systematic uncertainties. For both of these, the assumption of Gaussian statistics may be problematic, depending on the circumstances. For statistical effects, Poissonian rather than Gaussian statistics may apply to low-event-rate data, while, for systematic effects, lognormal statistics may be more appropriate than Gaussian statistics. The differences may or may not be small. But since the overarching goal is to provide a reaction rate with the most reliable uncertainty, it is crucial to analyze the D(p,$\gamma$)$^3$He data using an independent method that does not make the implicit assumption of Gaussian statistics throughout.

The first step in this direction was made in the work of \citet{iliadis16} which analyzed D(p,$\gamma$)$^3$He data using a Bayesian hierarchical model \citep{2017bmad}. A Bayesian model is not restricted to Gaussian statistics, but allows for the implementation of any probability distributions that best describe the data at hand. Recently, such models have been applied successfully to estimate other BBN reaction rates \citep{gomez17,deSouza:2019gf,deSouza:2019gi,de_Souza_2020}. We will provide a discussion of our Bayesian model in Section~\ref{sec:Bayemodel}.

\subsection{Fitting functions} 
\label{sec:function}
Fitting D(p,$\gamma$)$^3$He S-factor data requires the user to adopt a choice for the theoretical relationship between S-factor and energy. In many previous works, e.g., \citet{Mossa20,Yeh21,Pis21}, a polynomial function was assumed for this relationship. Polynomial models have a number of advantages: a simple form, well-known properties, moderate flexibility of shapes, and computational ease of use. However, they also have limitations: poor interpolatory and extrapolatory properties, a poor trade-off between degree and shape, and a disregard of nuclear theory. In extreme cases, these issues could lead to numerically unstable models. 

A different strategy was employed by \citet{Coc15,iliadis16}, who adopted, for the theoretical S-factor, the predictions of a microscopic nuclear model. Their model had only a single open parameter, i.e., a scale factor by which nuclear theory was multiplied during the fitting process. Their reasoning was that, while current microscopic nuclear models may be expected to reliably reproduce the energy dependence of the S-factor, the absolute normalization should be determined by the experimental data. This method has a number of advantages over the use of polynomials: better interpolatory and extrapolatory properties, and fewer fit parameters resulting in more stable models. The disadvantages are that the nuclear theoretical S-factor cannot be written as an analytical expression anymore, but must be computed numerically, and that a specific microscopic model may deviate from the true S-factor not only in scale, but also in energy dependence.

Note that we are not claiming that one method is inherently better than the other. Since the goal is to predict the most reliable S-factor based on data, it is important to investigate any differences that are obtained with different theoretical relationships between S-factor and energy. Therefore, we will construct Bayesian models for both a polynomial and a microscopic theory prescription. In the first case, we follow \citet{Mossa20,Yeh21,Pis21} and assume a cubic polynomial
\begin{equation}
\label{eq:poly}
S_{true}(E) = S_0 + S_1 E + S_2 E^2 + S_3 E^3,
\end{equation}
implying four fit parameters, $S_0$, $S_1$, $S_2$, and $S_3$. 

For the second case, we will modify the procedure used in \citet{Coc15,iliadis16}. Microscopic cross section calculations represent model-based Hamiltonian approaches with {\it a priori} difficult-to-quantify uncertainties. Bias in the microscopic theory S-factor may arise from the truncation of the set of states in the basis used to determine the matrix elements, or the exclusion of operators in the Hamiltonian. These issues cannot be entirely disregarded, despite the fact that these types of model calculations are usually tuned to experimental scattering data and binding energies. To allow for the possibility of bias in the microscopic theory S-factor, we will introduce two fit parameters instead of one; a multiplicative scale factor, $a$, by which nuclear theory, $S_{nuc}$, is multiplied, similar to \citet{Coc15,iliadis16}, and an offset, $b$, 
\begin{equation}
\label{eq:theory}
S_{true}(E) = a S_{nuc}(E) + b.
\end{equation}
Our results will be of interest not only for the D(p,$\gamma$)$^3$He reaction rate, but also to assess the microscopic model S-factor prediction (i.e., by how much the fit results differ from $a$ $=$ $1$ and $b$ $=$ $0$).

For $S_{nuc}$ we adopt the numerical values of \citet{marcucci05}, although more recent results have been published by the same group \citep{Mar16}. The absolute magnitudes of the two theoretical predictions differ by $>$7\% near $91$~keV (i.e., the center of the Gamow peak for $0.8$~GK). However, we are only interested in the energy dependence, which agrees within $0.8$\% over the energy range of interest for BBN. We have verified that using one microscopic result over the other will give rise to different values of the parameters $a$ and $b$, but the fitted S-factors agree within uncertainties. 

\section{Bayesian model} 
\label{sec:Bayemodel}
The hierarchical Bayesian model in the present work is similar to the one in \citet{iliadis16,de_Souza_2020}. This model allows us to take all relevant effects and processes into account impacting the measured data. The framework for Bayesian Inference rests on Bayes' Theorem \citep{2017bmad}
\begin{equation}
\label{eq:Bayes}
    p(\theta|y) = \frac{\mathcal{L}(y|\theta)\pi(\theta)}{\int \mathcal{L}(y|\theta)\pi(\theta)d\theta},
\end{equation}
where the data are represented by the vector $y$ and the complete set of model parameters is described by the vector $\theta$. All factors in Equation (\ref{eq:Bayes}) serve as probability densities: $\mathcal{L}(y|\theta)$ is the likelihood, i.e., the probability that the data, $y$, were obtained assuming given values of the model parameters, $\theta$; $\pi(\theta)$ is the prior, which represents our state of knowledge about each parameter before seeing the data; the product of the likelihood and the prior defines the posterior, $p(\theta|y)$, i.e., the probability of obtaining the values of a specific set of model parameters given the data. The posterior represents an update to our prior state of knowledge about the model parameters once new data becomes available. The denominator, termed the evidence, is a normalization factor and is not pertinent toward the discussion of the present work. 

The astrophysical S-factor is related to the reaction cross section by $S(E) \equiv E e^{2 \pi \eta} \sigma(E)$, where $\eta$ is the Sommerfeld parameter. When the experimental S-factor is subject to statistical uncertainties only, $\sigma_{stat}$, the likelihood is given by\footnote{\bf{
In \citet{iliadis16,gomez17}, the statistical uncertainties were described by lognormal probability densities. Here, we will follow later work \citep{deSouza:2019gf,deSouza:2019gi,de_Souza_2020} that employed normal densities because the statistical uncertainties are relatively small such that the difference between normal and lognormal densities becomes insignificant.
}
}
\begin{equation}
\label{eq:likelihood} 
    \mathcal{L}(S^{exp}|\theta)=\prod_{i=1}^N \frac{1}{\sigma_{stat,i}\sqrt{2\pi}}
    e^{-\frac{[S_i^{exp} - S(\theta)_i]^2}{2\sigma^2_{stat,i}}},
\end{equation}
where $S(\theta)_i$ is the theoretical S-factor, given by Equation~(\ref{eq:poly}) or (\ref{eq:theory}), and the product runs over all the data points, labeled by $i$. This likelihood represents a product of normal distributions, each with a mean of $S(\theta)_i$ and a standard deviation of $\sigma_{stat,i}$, given by the experimental statistical uncertainty of datum $i$. In symbolic notation, we write the above expression as
\begin{equation}
\label{eq:Si}
S_i^{exp} \sim \mathrm{Normal}(S(\theta)_i,\sigma^2_{stat,i}),
\end{equation}
where ``$\mathrm{Normal}$'' denotes a normal (Gaussian) probability density and the symbol ``$\sim$'' means ``has the probability distribution of.''

In many cases, the observed scatter of the measured data cannot be explained solely by the reported statistical uncertainties, suggesting the presence of additional sources of statistical uncertainties unknown to the experimenter. We utilize the expression \emph{extrinsic uncertainty} for describing such effects \citep{deSouza:2019gf}. Since the observed scatter in the data points contains the information on any additional statistical uncertainties, our model can predict the magnitude of the extrinsic uncertainty for each data set. When both statistical and extrinsic uncertainties are present in a measurement, the overall likelihood is given by a nested (hierarchical) expression. Using symbolic notation, we can write
\begin{equation}
\label{eq:Si2}
 S^{\prime}_{i} \sim \mathrm{Normal}(S(\theta)_{i},\sigma^2_{extr}),
\end{equation}
\begin{equation}
\label{eq:Si3}
 S^{exp}_i \sim \mathrm{Normal}(S^{\prime}_i,\sigma^2_{stat;i}).
\end{equation}
Equations~(\ref{eq:Si2}) and (\ref{eq:Si3}) provide an intuitive approach for constructing the overall likelihood: first, statistical uncertainties, quantified by the standard deviation, $\sigma_{extr}$, of a normal probability density, perturb the true (but unknown) value of the S-factor at the given energy of a data point $i$, $S(\theta)_i$, to produce a value of $S^{\prime}_i$; second, the latter value is perturbed, in turn, by the reported experimental statistical uncertainty, quantified by the standard deviation, $\sigma_{stat;i}$, of a normal probability density, to produce the measured value of $S^{exp}_i$. The above example demonstrates how experimental effects impacting the data can be implemented into a Bayesian model.
 
Each of the model parameters contained in the vector $\theta$ requires a prior. Priors are chosen to best represent the physics involved. For example, if our model includes a parameter, $A$, and all we know is that its value lies somewhere in a region from zero to $A_{max}$, we can write a prior as
\begin{equation}
\label{eq:prior}
 A \sim \mathrm{Uniform}(0,A_{max}),
\end{equation}
where ``$\mathrm{Uniform}$'' denotes a uniform probability density. Instead of uniform priors, we will be using, for most parameters, normal probability densities with a large spread. These are non-informative, normalizable, and, unlike uniform priors, do not have sharp (unphysical) boundaries.
 
As mentioned in Section~\ref{sec:selection} (see also Table 1), we divided the D(d,$\gamma$)$^3$He experiments into two groups, i.e., those with reliable absolute cross section normalizations (``absolute data''), and those for which we only utilize the energy dependence of the cross section (``relative data''). We discuss below the prior choices for these two groups in more detail. 

For the first group \citep{War63,Ma97,Sch97,Cas02,Tisma:2019ug,Mossa20,Turkat21}, we adopted reported systematic uncertainties and included them as normalization factors into our model. For example, a systematic uncertainty of, say, $\pm5\%$, implies that the systematic factor uncertainty is 1.05. The true value of the normalization factor, $f$, is unknown at this stage. However, we do know that the expectation value of the normalization factor is unity. If not for this, we would have corrected the data for the systematic effect. A useful distribution we employ for normalization factors is the lognormal probability density, which is characterized by two quantities: the location parameter, $\mu$, and the spread parameter, $\sigma$. The median value of the lognormal distribution is given by $x_{med} = e^{\mu}$, while the factor uncertainty, for a coverage probability of $68\%$, is $f.u. = e^{\sigma}$. We include in our Bayesian model a systematic effect on the S-factor as an informative, lognormal prior with a median of $x_{med} = 1.0$ (or $\mu = \ln x_{med}=0$), and a factor uncertainty given by the systematic uncertainty; i.e., in the above example, $f.u. = 1.05$ (or $\sigma = \ln f.u. = \ln (1.05)$). The prior is then explicitly given by 
\begin{equation}
\label{eq:lognor}
 \pi(f) = \frac{1}{\ln (f.u.) \sqrt{2\pi}f}e^{-\frac{[\ln f]^2}{2[\ln (f.u.)]^2}},
\end{equation}
or
\begin{equation}
\label{eq:lognor2}
f \sim \mathrm{LogNormal}(0, [\ln (f.u.)]^2),
\end{equation}
where ``$\mathrm{LogNormal}$'' denotes a lognormal probability density. For more information on this choice of prior, see \citet{iliadis16}. For the six absolute data sets, we adopted for the factor uncertainties, $f.u.$, the reported values of 1.14 \citep{Turkat21}, 1.027 \citep{Mossa20}, 1.10 \citep{Tisma:2019ug}, 1.045 \citep{Cas02}, 1.09 \citep{Sch97}, 1.09 \citep{Ma97}, and 1.10 \citep{War63}. More details are provided in Appendix~\ref{sec:app}.

 
For the relative data \citep{Gri55,Gri62,Gri63,Bai70}, we chose to scale the true (unknown) S-factor by a factor of $10^g$, with a non-informative prior of
\begin{equation}
\label{eq:reldata}
 g \sim \mathrm{Uniform}(-1,1),
\end{equation}
corresponding to a uniform prior between $-1$ and $+1$. In other words, the normalization factor, $10^g$, is varied by up to one order of magnitude up or down during the sampling. Therefore, the relative data provide only information on the energy dependence of the S-factor, but none on the absolute normalization. Furthermore, it is not possible to extract reliable statistical uncertainties for these data, as is explained in more detail in Appendix~\ref{sec:app}. Therefore, they will be subject only to extrinsic uncertainties, which are included in the likelihood according to Equation~(\ref{eq:Si2}).
 
For both absolute and relative data, the extrinsic uncertainties of the measured cross sections are inherently unknown to the experimenter. Thus, we will adopt broad normal priors that are truncated at zero, with a standard deviation of $10^{-4}$~MeVb. 
 
Our complete Bayesian model is summarized below in symbolic notation:
\begin{alignteo}
\label{eq:model}
    & \textrm{{\bf Model relationships:}} \notag \\
    & \mathrm{(1)}~ S_i(E) = a S_{nuc}(E) + b \notag\\
    & \mathrm{(2)}~ S_i(E) = S_0 + S_1 E + S_2 E^2 + S_3 E^3 \notag\\
    & \textrm{{\bf Parameters:}} \notag \\
    & \mathrm{(1)}~ \theta \equiv (a, b) \notag \\  
    & \mathrm{(2)}~ \theta \equiv (S_0, S_1, S_2, S_3) \notag \\  
    & \textrm{{\bf Likelihood (absolute data):}}\\ 
    & S^{\prime}_{i,k} = f_{k} \times S_i, \notag \\
    & S^{\prime\prime}_{i,k}  \sim \rm{Normal}(S^{\prime}_{i,k}, \sigma_{extr}^2), \notag \\   
    & S^{exp}_{i,k}  \sim \rm{Normal}(S^{\prime\prime}_{i,k}, \sigma_{stat,i}^2). \notag \\
    & \textrm{{\bf Likelihood (relative data):}} \notag \\  
    & S^{\prime}_{i,j} = 10^{g_{j}} \times S_i, \notag \\
    & S^{exp}_{i,j}  \sim \rm{Normal}(S^{\prime}_{i,j}, \sigma_{extr}^2), \notag \\
    & \textrm{{\bf Priors:}}\notag \\ 
    & \sigma_{\mathrm{extr}} \sim \rm{TruncNormal}(0, [10^{-4}~\mathrm{MeVb}]^2), \notag \\ 
    & f_{k} \sim \mathrm{LogNormal}(0, [\ln(f.u.)_k]^2),  \notag \\
    & g_{j} \sim \rm{Uniform}(-1,+1),  \notag  
\end{alignteo}
The index $i$ labels individual data points, $j = 1,...,4$ denotes the relative data sets with information on the energy dependence only, and $k$ $=$ $1,...,6$ labels the absolute data sets that include information on the cross section normalizations. The ``TruncNormal()'' prior refers to a truncated normal probability distribution; i.e., a density with a maximum at zero that excludes negative values.  
 
For the analysis of Bayesian models, we employ the program JAGS (“Just Another Gibbs Sampler”) using Markov chain Monte Carlo (MCMC) sampling \citep{Plummer03jags:a}. Specifically, we will employ the rjags package that works directly with JAGS within the R language \citep{Rcitation}. Running a JAGS model refers to generating random samples from the posterior distribution of model parameters. This involves the definition of the model, likelihood, and priors, as well as the initialization, adaptation, and monitoring of the Markov chain.
 
\section{Results} 
\label{sec:results}


The MCMC sampling will provide the posteriors of all parameters. We computed three MCMC chains, where each had a length of $10^6$ steps after disregarding the burn-in samples ($3\times 10^5$ steps for each chain). This ensured that the chains reached equilibrium and that Monte Carlo fluctuations were negligible compared to statistical, systematic, and extrinsic uncertainties.

\subsection{S-factors and physical model parameters} 
\label{sec:sfac}
Values of the physical model parameters are listed in Table~\ref{tab:results}. Our best-fit S-factors using the two-parameter (nuclear theory) and four-parameter (polynomial) fit functions (Section~\ref{sec:function}) are presented in the top panels of Figures~\ref{fig:sfactor} and \ref{fig:sfactor2}, respectively. The bottom panels show the S-factor residuals of the fit and the data. The gray-shaded region in the top panels depicts the Gamow peak for a temperature of $T$ $=$ $0.8$~GK. The maximum of the peak occurs at $91$~keV in the $d$ $+$ $p$ center of mass, which is a useful energy for S-factor comparisons. 

\begin{figure}
\includegraphics[width=1\linewidth]{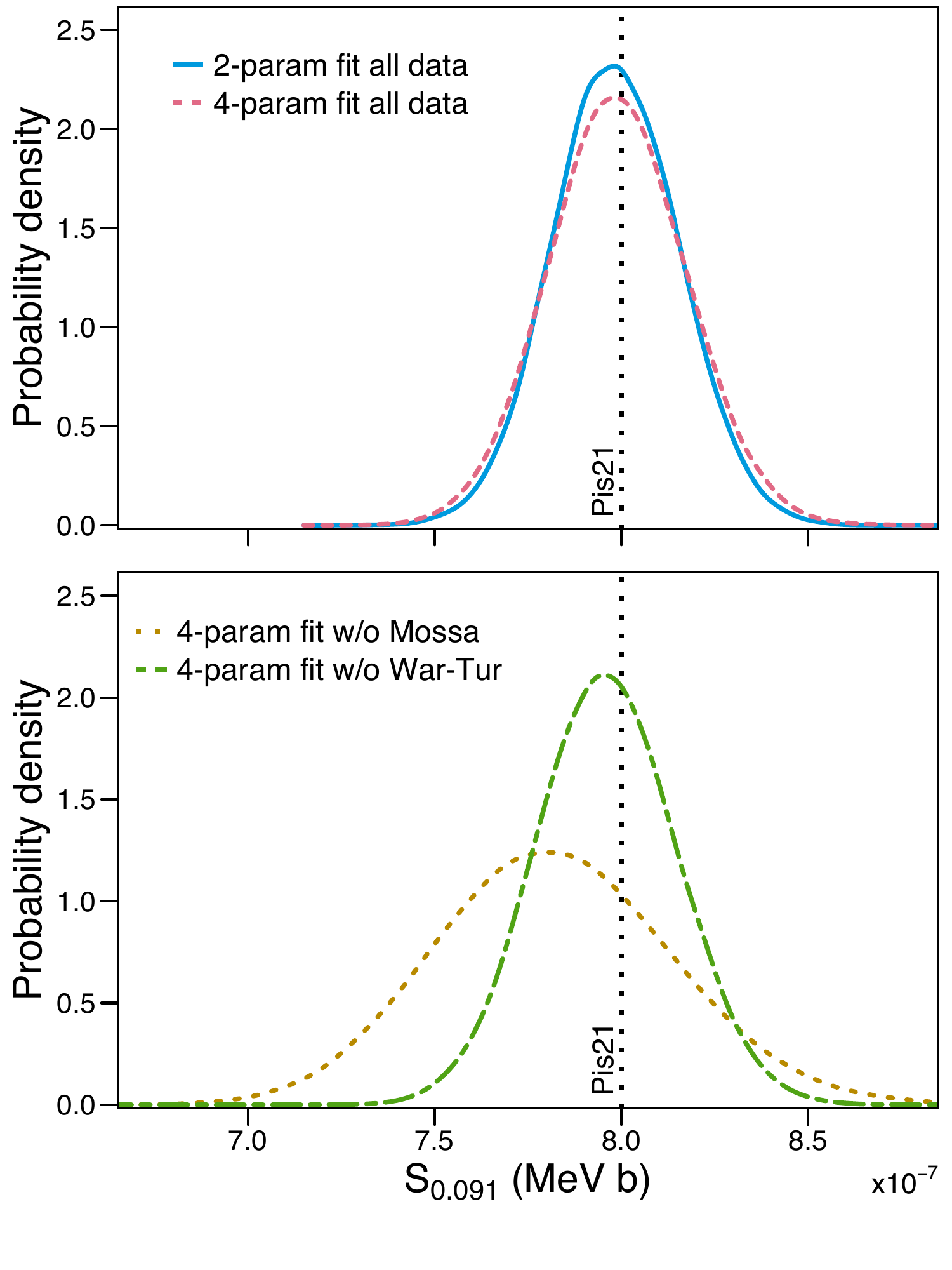}
\caption{Marginalized posteriors of the S-factor at a representative center-of-mass energy of $91$~keV; i.e., at the center of the Gamow peak for a temperature of $T$ $=$ $0.8$~GK. (Top) Posteriors obtained with our two-parameter (solid blue) and four-parameter model (dashed red) Bayesian fit (Section~\ref{sec:function}), including all eleven data sets listed with check marks in column 2 of Table~\ref{tab:references}. The solid blue and dashed red lines represent vertical slices at $91$~keV of the fits shown in Figures~\ref{fig:sfactor} and \ref{fig:sfactor2}, respectively. For comparison, the vertical black dotted line depicts the mean value for the polynomial fit of \citet{Pis21}. (Bottom) The dotted orange line corresponds to the posterior of our four-parameter model fit obtained by excluding the data measured by \citet{Mossa20}, whereas the dashed-dotted green line indicates the posterior obtained by excluding the data measured by \citet{War63,Turkat21}.} 
\label{fig:sfac91}
\end{figure}

For the two-parameter fit we find a value of $S_{2p}$(91~keV) $=$ $(7.98\pm0.17)$ $\times$ $10^{-7}$~MeVb, where the uncertainties were derived from the 16, 50, and 84 percentiles (Table~\ref{tab:results}). The four-parameter (polynomial) fit yields $S_{4p}$(91~keV) $=$ $(7.99\pm0.18)$ $\times$ $10^{-7}$~MeVb, consistent with the two-parameter fit. The marginalized posteriors from which these results were derived are depicted in the top panel of Figure~\ref{fig:sfac91}. The solid blue and dashed red lines correspond to our two-parameter and four-parameter fits, respectively. For comparison, the vertical black dotted line indicates the mean value of the four-parameter fit result of \citet{Pis21}, which corresponds to $S_{prev}$(91~keV) $=$ $8.00$ $\times$ 10$^{-7}$~MeVb (Table~\ref{tab:results}). It can be seen that their mean value is near the center of the posterior densities derived in the present work. \citet{Pis21} did not directly report their S-factor uncertainties and, therefore, we cannot compare our results to theirs (see Appendix~\ref{sec:app2}). Numerical values of the parameters for the present and previous fits are also listed in Table~\ref{tab:results}.


%
\begin{deluxetable*}{lccc}
\label{tab:results} 
\tablecaption{Fit parameters for the physical model and S-factor prediction at $91$~keV.}
\tablewidth{\linewidth}
\tablehead{
Parameter  &  Present &  Present &  Previous\tablenotemark{c}  \\
           & 2-param\tablenotemark{a,e}  &  4-param\tablenotemark{b,e}  &  4-param\tablenotemark{b}
} 
\startdata
Scale factor            \tablenotemark{a}   &  0.953$\pm$0.024                          &                                                   &       \\ 
Offset (MeVb)           \tablenotemark{a}   & (2.0$\pm1.0$)$\times$10$^{-8}$            &                                                   &  \\ 
$S_0$ (MeVb)            \tablenotemark{b}   &                                           &  (2.19$\pm$0.10)$\times$10$^{-7}$                 &   2.12$\times$10$^{-7}$   \\
$S_1$ (b)               \tablenotemark{b}   &                                           &  (5.80$\pm$0.24)$\times$10$^{-6}$                 &   5.97$\times$10$^{-6}$    \\
$S_2$ (b MeV$^{-1}$)    \tablenotemark{b}   &                                           &  (6.34$_{-0.82}^{+0.88}$)$\times$10$^{-6}$        &   5.46$\times$10$^{-6}$    \\
$S_3$ (b MeV$^{-2}$)    \tablenotemark{b}   &                                           &  ($-$2.20$\pm$0.52)$\times$10$^{-6}$              &   $-$1.66$\times$10$^{-6}$ \\
\hline
$S$(91~keV) (MeVb) \tablenotemark{d}        &  (7.98$\pm$0.17)$\times$10$^{-7}$         &  (7.99$\pm$0.18)$\times$10$^{-7}$                 &   8.00$\times$10$^{-7}$        \\
\enddata
\tablenotetext{a}{Fit parameters according to $S(E)$ $=$ (Scale factor) $\times$ $S_{theory}$ $+$ (Offset), where $S_{theory}$ denotes the theoretical S-factor of \citet{marcucci05}.}
\tablenotetext{b}{Coefficients of S-factor parameterization, according to $S(E)$ $=$ $S_0$ $+$ $S_1 E$ $+$ $S_2 E^2$ $+$ $S_3 E^3$ MeV b, with the center-of-mass energies given in MeV.}
\tablenotetext{c}{From \citet{Pis21}, as indicated by the expression given in their appendix; no uncertainties have been reported in their work (see Appendix~\ref{sec:app2}).}
\tablenotetext{d}{Astrophysical S-factor at $91$~keV, i.e., the center of the Gamow peak for $0.8$~GK.}
\tablenotetext{e}{Uncertainties and median values derived from 16, 50, and 84 percentiles.}
\end{deluxetable*}
%

\begin{figure}
\includegraphics[width=0.9\linewidth]{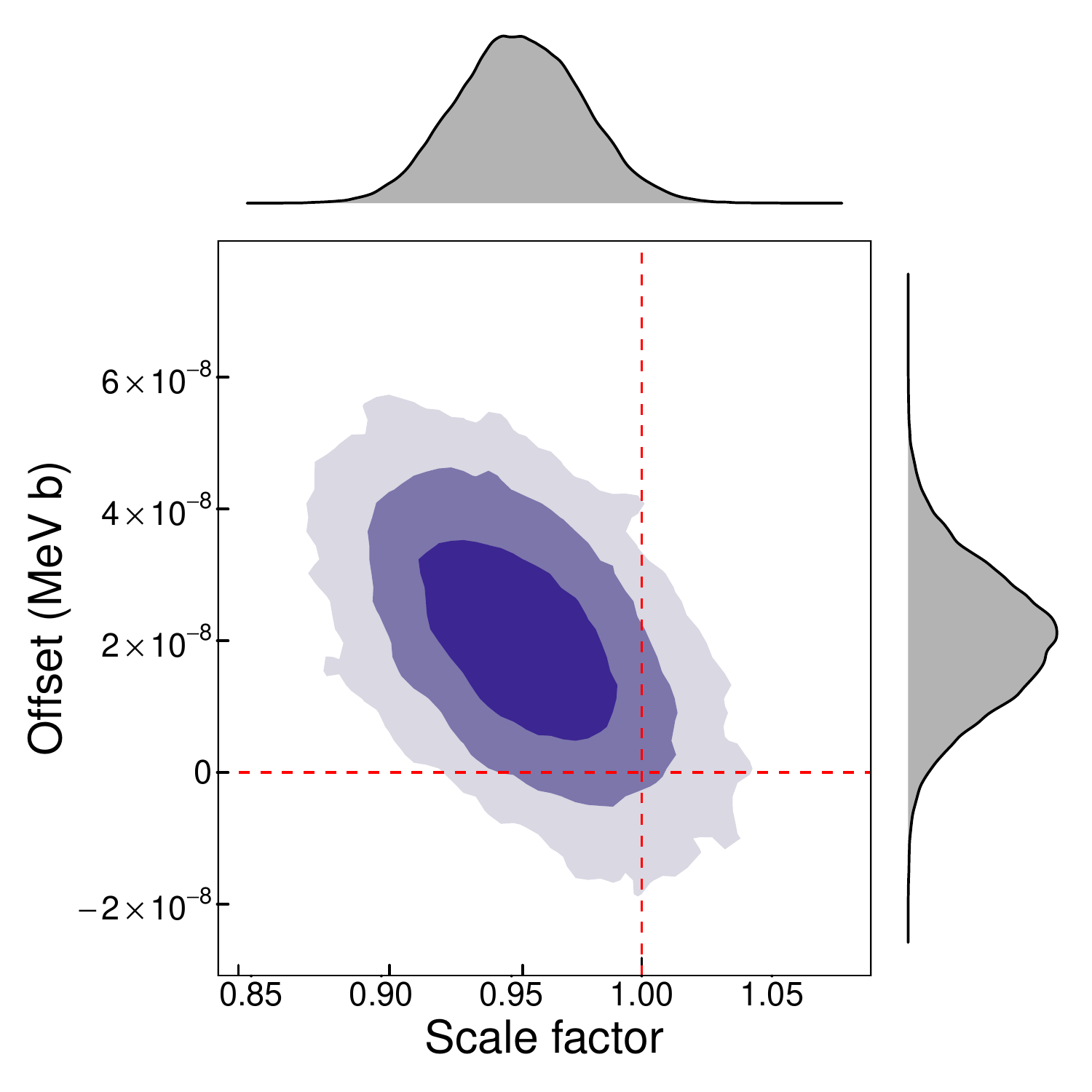}
\caption{One- and two-dimensional projections of the posterior probability distributions for the two physical model parameters of our two-parameter fit (Section~\ref{sec:results}). The gray areas at the top and right depict the one-dimensional marginalized posteriors. The central panel indicates the 68\%, 95\%, and 99.7\% credible intervals (dark, medium, and light blue, respectively), of the two-dimensional pairwise posterior projections. The red dashed cross hairs correspond to a scale factor of unity ($a$ $=$ $1$) and an offset of zero ($b$ $=$ $0$). See Equation~(\ref{eq:theory}).
}
\label{fig:2D}
\end{figure}

Regarding our two-parameter fit, recall from Section~\ref{sec:function} that the values $a$ $=$ $1$ and $b$ $=$ $0$ correspond to a microscopic theory S-factor consistent with the data. One- and two-dimensional projections of the posterior probability distributions of the two parameters are depicted in Figure~\ref{fig:2D}. The scale factor, $a$ $=$ $0.953\pm0.024$, differs significantly from unity, and the offset, $b$ $=$ (2.2$\pm1.0$)$\times$10$^{-8}$, differs from zero (see red dashed cross hairs in the central panel of Figure~\ref{fig:2D}). Thus, it appears that both the absolute magnitude and the energy dependence of the S-factor predicted by the microscopic model of \citet{marcucci05} are in conflict with the available data.

\subsection{Data set parameters} 
\label{sec:datapar}
\begin{figure}
\includegraphics[width=0.9\linewidth]{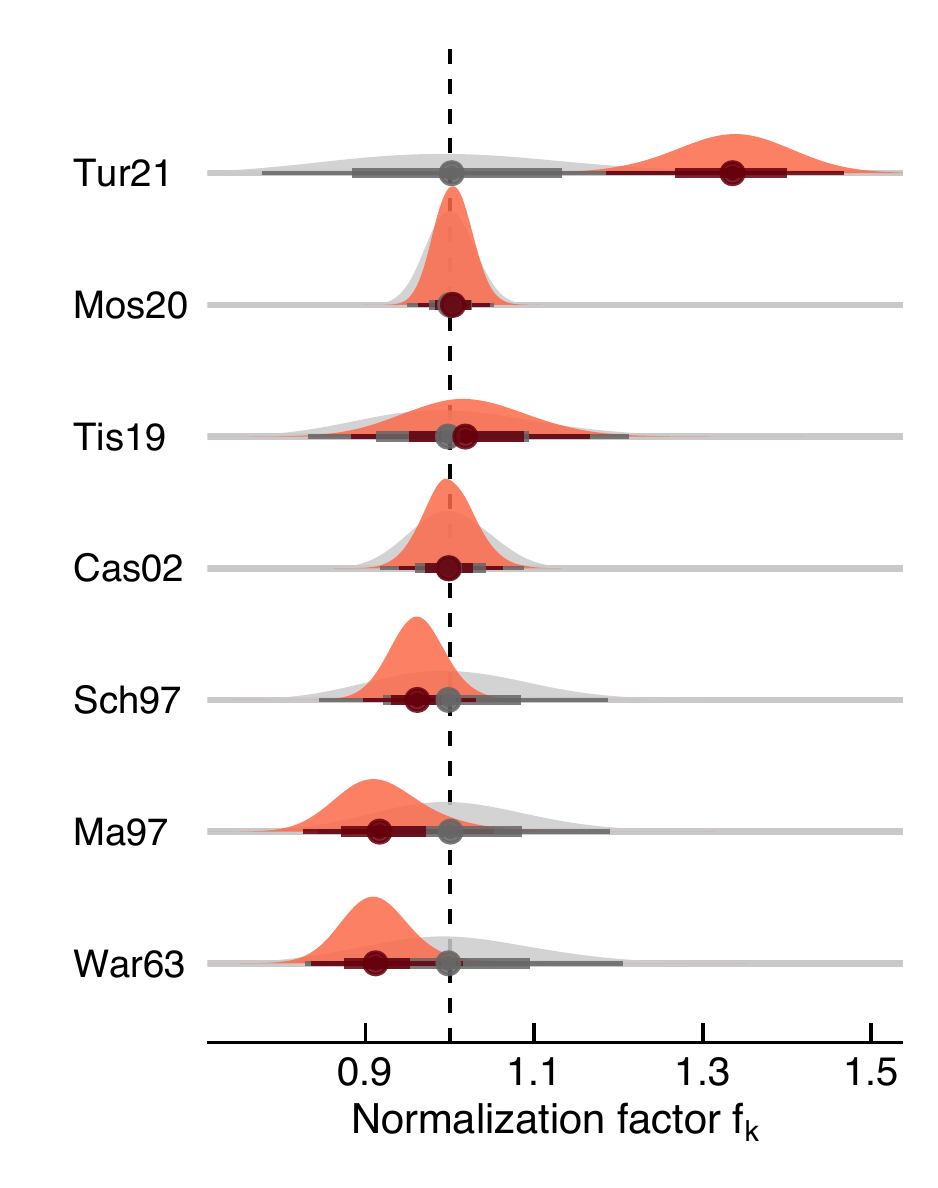}
\caption{Marginalized posteriors (red) of the normalization factors, $f_k$, for each of the seven experiments that reported absolute S-factors. The results are obtained with the present two-parameter fit displayed in Figure~\ref{fig:sfactor}. 
The full circles, horizontal thick bars, and horizontal thin bars below each distribution show the median, and the 68\% and 95\% credible intervals, respectively. 
For comparison, the corresponding prior distributions are displayed in gray. 
For the data set labels, see Figure~\ref{fig:sfactor}.
}
\label{fig:post2}
\end{figure}

Posteriors of the normalization factors, $f_k$, for each of the experiments that reported absolute S-factors are displayed as red areas in Figure~\ref{fig:post2}. These results were obtained with the present two-parameter fit given in Figure~\ref{fig:sfactor}. For comparison, the densities shown in gray depict the priors, with their spreads determined by the reported systematic uncertainties (see Section~\ref{sec:Bayemodel}). Numerical values are listed in Table~\ref{tab:modpar}, indicating agreement between the results of the two- and four-parameter fits. Recall that these values represent factors by which the true cross section is multiplied to agree with the data, as explained in Section~\ref{sec:Bayemodel}. It can be seen that, with one exception, the priors and posteriors overlap significantly, meaning that these works appear to have reported reliable systematic uncertainties. The exception is the data set of \citet{Turkat21} with a normalization factor of $f_1$ $\approx$ $1.3$, which significantly exceeds their reported systematic uncertainty of $\approx$14\%. This implies that our best fit line is pulled only weakly towards this particular data set compared to the other data. 

\begin{deluxetable*}{lll}
\label{tab:modpar} 
\tablecaption{Fit parameters for modeling the data sets.}
\tablewidth{\linewidth}
\tablehead{
Parameter  &  Present &  Present   \\
           & 2-param\tablenotemark{a}  &  4-param\tablenotemark{b}  
} 
\startdata
$f_1$ \tablenotemark{c}                                 &     1.335$\pm$0.069                                  &  1.317$\pm$0.070               \\
$f_2$ \tablenotemark{c}                                 &     1.004$\pm$0.022                                  &  1.003$\pm$0.023                 \\
$f_3$ \tablenotemark{c}                                 &     1.018$\pm$0.072                                  &  1.019$\pm$0.071                 \\
$f_4$ \tablenotemark{c}                                 &     0.996$\pm$0.030                                  &  0.993$\pm$0.032                 \\
$f_5$ \tablenotemark{c}                                 &     0.962$\pm$0.033                                  &  0.955$\pm$0.033                 \\
$f_6$ \tablenotemark{c}                                 &     0.917$_{-0.048}^{+0.058}$                        &  0.917$_{-0.048}^{+0.058}$        \\
$f_7$ \tablenotemark{c}                                 &     0.912$\pm$0.041                                  &  0.956$\pm$0.048                 \\ 
$\sigma^{\mathrm{extr}}_1$ (MeVb) \tablenotemark{d}     &     (5.9$_{-2.2}^{+2.9}$)$\times$10$^{-7}$        &  (6.0$_{-2.2}^{+2.9}$)$\times$10$^{-7}$   \\                       
$\sigma^{\mathrm{extr}}_2$ (MeVb) \tablenotemark{d}     &     (2.00$_{-0.43}^{+0.60}$)$\times$10$^{-8}$        &  (2.14$_{-0.47}^{+0.66}$)$\times$10$^{-8}$   \\                       
$\sigma^{\mathrm{extr}}_3$ (MeVb) \tablenotemark{d}     &     $\leq$6.0$\times$10$^{-7}$                       &  $\leq$6.0$\times$10$^{-7}$   \\                       
$\sigma^{\mathrm{extr}}_4$ (MeVb) \tablenotemark{d}     &     $\leq$9.8$\times$10$^{-9}$                       &  $\leq$9.8$\times$10$^{-9}$   \\                       
$\sigma^{\mathrm{extr}}_5$ (MeVb) \tablenotemark{d}     &     (2.56$_{-0.72}^{+1.20}$)$\times$10$^{-8}$        &  (2.56$_{-0.73}^{+1.21}$)$\times$10$^{-8}$   \\                       
$\sigma^{\mathrm{extr}}_6$ (MeVb) \tablenotemark{d}     &     $\leq$4.3$\times$10$^{-7}$                       &  $\leq$4.3$\times$10$^{-7}$   \\                       
$\sigma^{\mathrm{extr}}_7$ (MeVb) \tablenotemark{d}     &     $\leq$3.4$\times$10$^{-6}$                       &  $\leq$2.2$\times$10$^{-6}$   \\                       
$\sigma^{\mathrm{extr}}_8$ (MeVb) \tablenotemark{d}     &     (1.14$_{-0.22}^{+0.32}$)$\times$10$^{-7}$        &  (1.17$_{-0.23}^{+0.34}$)$\times$10$^{-7}$   \\                       
$\sigma^{\mathrm{extr}}_9$ (MeVb) \tablenotemark{d}     &     (4.00$_{-0.77}^{+1.13}$)$\times$10$^{-8}$        &  (4.00$_{-0.77}^{+1.13}$)$\times$10$^{-8}$   \\                       
$\sigma^{\mathrm{extr}}_{10}$ (MeVb) \tablenotemark{d}  &     $\leq$9.2$\times$10$^{-7}$                       &  $\leq$9.2$\times$10$^{-7}$   \\                       
$\sigma^{\mathrm{extr}}_{11}$ (MeVb) \tablenotemark{d}  &     (3.70$_{-0.65}^{+0.93}$)$\times$10$^{-7}$        &  (3.42$_{-0.61}^{+0.86}$)$\times$10$^{-7}$   \\                       
\enddata
\tablenotetext{a}{Fit parameters according to $S(E)$ $=$ (Scale factor) $\times$ $S_{theory}$ $+$ (Offset), where $S_{theory}$ denotes the theoretical S-factor of \citet{marcucci05}.}
\tablenotetext{b}{Coefficients of S-factor parameterization, according to $S(E)$ $=$ $S_0$ $+$ $S_1 E$ $+$ $S_2 E^2$ $+$ $S_3 E^3$ MeV b.}
\tablenotetext{c}{Normalization factor. Subscripts refer to the data of: (1) \citet{Turkat21}; (2) \citet{Mossa20}; (3) \citet{Tisma:2019ug}; (4) \citet{Cas02}; (5) \citet{Sch97}; (6) \citet{Ma97}; (7) \citet{War63}; (8) \citet{Bai70}; (9) \citet{Gri63}; (10) \citet{Gri62}; (11) \citet{Gri55}.}
\tablenotetext{d}{Extrinsic uncertainty, assumed to be caused by an additional (unknown) source of statistical scatter. Upper limit values correspond to 97.5 percentiles of the posteriors.}
\end{deluxetable*}

The spread parameters of the extrinsic scatter, $\sigma_k^{extr}$, are also listed in Table~\ref{tab:modpar} for both our two- and four-parameter fits. The values from both models are in agreement, and indicate that any unreported additional statistical scatter is either smaller or comparable in magnitude to the reported statistical uncertainties. The only exception is the data set of \citet{Sch97}, for which we obtain an extrinsic scatter ($\sigma_4^{extr}$ $\approx$ $2.6\times 10^{-8}$~MeVb) that significantly exceeds their reported statistical uncertainties ($\approx$ $1.0 \times 10^{-8}$~MeVb).

\subsection{Tests} 
\label{sec:resultstests}
To assess how robust our S-factor fits are to the data selection, we repeated the analysis by disregarding specific data sets. In the following, we will focus on the results obtained using the four-parameter fit model. Although we will quote for these tests S-factor values at a representative center-of-mass energy of $91$~keV only (i.e., at the center of the Gamow peak for a temperature of $T$ $=$ $0.8$~GK), it must be emphasized that deuterium BBN takes place over a range of energies and that quantitatively different results are obtained for different energies. 

For the first test, we disregarded the data measured by \citet{Mossa20}, which cover center-of-mass energies between $0.03$~MeV and $0.3$~MeV. The result is $S_{4p}(91~\mathrm{keV})$ $=$ $(7.81\pm0.31)$ $\times$ $10^{-7}$~MeVb. The corresponding probability density is depicted as dotted orange line in the lower panel of Figure~\ref{fig:sfac91}. Disregarding the data of \citet{Mossa20} not only lowers the mean S-factor at $91$~keV by 2.3\%, but also increases its uncertainty (from 2.2\% to 4.0\%), which can be seen by comparing the orange-dotted line to the blue or red lines in the top panel of Figure~\ref{fig:sfac91}. The reason is that this data set has the smallest overall uncertainties (see Appendix~\ref{sec:luna}) among all the data we took into account (Table~\ref{tab:results}). Therefore, it pulls on the best-fit S-factor more strongly than the other data.

For the second test, we disregarded the data of \citet{War63,Turkat21}. These two experiments provided the only absolute S-factors in the higher energy range ($\approx$0.3 $-$ $1.6$~MeV). In this case, we obtained a value of $S_{4p}(91~\mathrm{keV})$ $=$ $(7.96\pm0.18)$ $\times$ $10^{-7}$~MeVb. The probability density, shown as dashed-dotted green line in the lower panel of Figure~\ref{fig:sfac91}, is very close to the blue or red line in the top panel that take all data into account. This result confirms that the data at higher energies have only a small impact on the predicted S-factor near an energy of $91$~keV. 

Finally, we disregarded all relative data sets, i.e., those with a superscript ``e'' in column 2 of Table~\ref{tab:references}. The resulting S-factor value at $91$~keV was the same as that obtained for the full data set. This behavior is expected because, again, the best-fit S-factor is most strongly pulled towards the data of \citet{Mossa20} in this energy range. 

\section{Thermonuclear reaction rates} 
\label{sec:rates}
The thermonuclear reaction rate per particle pair, $N_A \langle \sigma v \rangle$, at a given plasma temperature, $T$, is given by \citep{Iliadis:2015ta}
\begin{equation}
\begin{split}
N_A \langle \sigma v \rangle & = \left(\frac{8}{\pi m_{01}}\right)^{1/2} \frac{N_A}{(kT)^{3/2}}  \\ 
&  \times \int_0^\infty e^{-2\pi\eta}\,S(E)\,e^{-E/kT}\,dE, 
\label{eq:rate}
\end{split}
\end{equation}
where $m_{01}$ is the reduced mass of projectile and target, $N_A$ is  Avogadro's constant, $k$ is the Boltzmann constant, and $E$ is the $d$ $+$ $p$ center-of-mass energy.

We computed the D(p,$\gamma$)$^3$He reaction rates by numerically integrating Equation~(\ref{eq:rate}). The S-factor is adopted from the samples of our  Bayesian fits (Section~\ref{sec:results}) and, thus, our values of $N_A \langle \sigma v \rangle$ fully contain the effects of statistical, systematic, and extrinsic uncertainties. We base these results on 75,000 random S-factor samples, which ensures that Markov chain Monte Carlo fluctuations are negligible compared to the reaction rate uncertainties. Our lower and upper integration limits were set to $10^{-6}$~MeV and $2.0$~MeV, respectively. Reaction rates are computed for $54$ different temperatures between $0.001$~GK and $4$~GK. At these temperatures, the data shown in Figure~\ref{fig:sfactor} fully cover the astrophysically important energy range. Numerical values of the reaction rates for our two-parameter function are listed in Table~\ref{tab:rate}. The recommended rates are computed from the 50th percentile of the probability density, while the factor uncertainty, $f.u.$, is obtained from the 16th and 84th percentiles \citep{Longland:2010is}. 

\begin{figure*}
\includegraphics[width=1\linewidth]{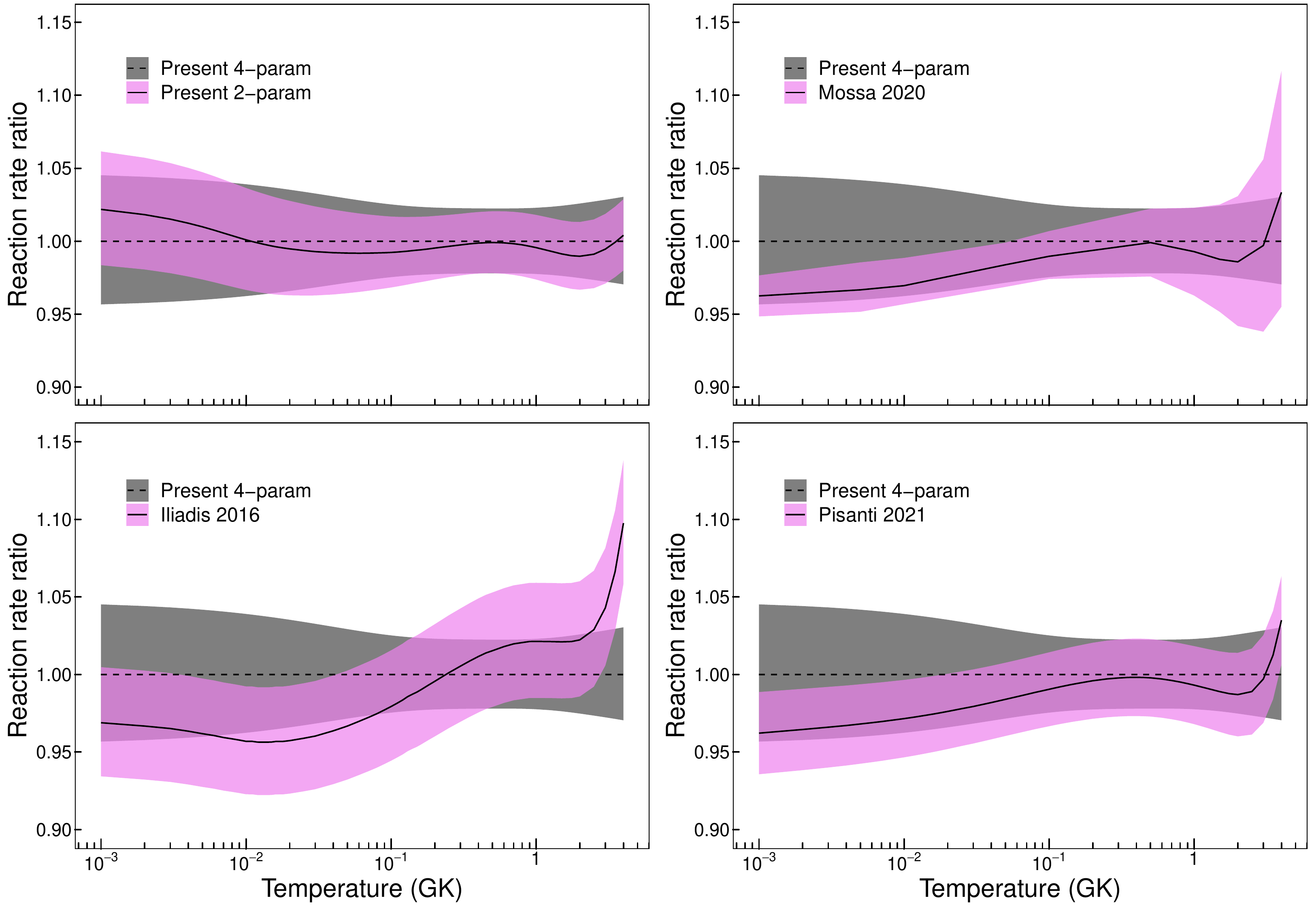}
\caption{Thermonuclear reaction rates of D(p,$\gamma$)$^3$He, normalized to the median (50th percentile) values of our four-parameter fit (Figure~\ref{fig:sfactor2}). These rates are shown in gray in all panels, while the results they are compared to are displayed in magenta. (Top left) Rates based on the present two-parameter fit results. (Bottom left) Rates of \citet{iliadis16}. (Top right) Rates of \citet{Mossa20}. (Bottom right) Rates of \citet{Pis21}. The present results and those of \citet{iliadis16} correspond to 68\% coverage probabilities, while the uncertainties of \citet{Mossa20,Pis21} have been obtained from a $\chi^2$ analysis. The most important temperature for deuterium BBN is near 0.8~GK.}
\label{fig:rates}
\end{figure*}

Reaction rates are displayed in Figure~\ref{fig:rates}. For better comparison, all rates have been divided by the median (50th percentile) values of our four-parameter fit (Figure~\ref{fig:sfactor2}). Our low (16th percentile) and high (84th percentile) rates are shown as gray bands centered around unity in all four panels. 

The top left panel compares the results for the two-parameter (shown in magenta) and four-parameter fits, both obtained in the present work. It can be seen that the rates agree within uncertainties. In other words, the choice of the particular fit function, Equation~(\ref{eq:poly}) or Equation~(\ref{eq:theory}), is inconsequential for the derived reaction rates. It is also apparent that the uncertainties of our four-parameter fit rates (gray) increase towards both the low and high temperature ends. This is exactly what one would expect from a polynomial fit function considering the lack of S-factor data beyond the low-energy and high-energy cutoffs. This effect, although less pronounced, is also visible in our two-parameter fit rates (magenta).

The bottom left panel compares our four-parameter fit results with those of the previous evaluation of \citet{iliadis16} (magenta), which has been used in the BBN simulations of \citet{Pitrou2018}. Although the gray and magenta bands overlap significantly, some differences are apparent. At the lowest temperatures, $T$ $=$ $0.001$~GK, the present recommended rates are higher by 3\% while, at $4$~GK, they are lower by 9\%. Also, the present rates have on average smaller uncertainties (between 2.1\% and 3.9\%, depending on temperature) compared to the 2016 rates (3.7\%). These differences are caused by the input information used to estimate the rates. First, only data up to center-of-mass energies of 200~keV were analyzed in \citet{iliadis16}, while here we are using all data up to 2~MeV (Table~\ref{tab:references}). Second, the data of \citet{Turkat21,Mossa20,Tisma:2019ug} were not available in 2016. 

In the top right panel, the magenta band depicts the rates of \citet{Mossa20}. At the low-temperature end, the previous rates have much smaller uncertainties ($1.5$\%) compared to our results ($3.9$\%), whereas at the high-temperature end their uncertainties ($8$\%) are much larger than ours ($2.5$\%). At the temperature most important for deuterium BBN nucleosynthesis, our uncertainties (2.2\%) are slightly smaller than those of \citet{Mossa20} (2.8\%). These differences are surprising considering that the results represented by both the gray and magenta bands were obtained by using the same fitting function; {\it i.e.,} a third-degree polynomial.

The bottom right panel presents the rates of \citet{Pis21} as a magenta band. These results largely overlap with our rates in the temperature range important for BBN, but a number of interesting differences are apparent. First, our recommended four-parameter fit rates are about 4\% larger at the lowest temperatures compared to the previous results. Furthermore, notice that the uncertainties of the magenta band do not increase towards the low- and high-temperature boundaries, which is surprising considering that the results are based on a polynomial fit function. By comparing the two magenta bands in the top right and bottom right panels, it can also be seen that the uncertainties of the rates by \citet{Mossa20} and \citet{Pis21} are quite different. This comparison calls into question the statement in \citet{Pis21} that ``...the small differences with respect to the expansion of the S-factor reported in [Mossa {\it et al.}] are due to the inclusion of the data of [Ti{\v s}ma {\it et al.}].'' We will comment in more detail on the S-factors of \citet{Mossa20,Pis21} in Appendix~\ref{sec:app2}.


\begin{deluxetable*}{cccccc}
\tablecaption{Present D(p,$\gamma$)$^3$He reaction rates.\tablenotemark{a}
\label{tab:rate}} 
\tablewidth{\columnwidth}
\tabletypesize{\footnotesize}
\tablecolumns{10}
\tablehead{
  \colhead{T (GK)} & \colhead{Rate}   & \colhead{$f.u.$} & \colhead{T (GK)} & \colhead{Rate}   & \colhead{$f.u.$} 
} 
\startdata
  0.001 & 1.454E-11 & 1.039  &   0.11    & 7.196E+00 & 1.024      \\
  0.002 & 2.007E-08 & 1.038  &   0.12    & 8.776E+00 & 1.024      \\
  0.003 & 6.495E-07 & 1.038  &   0.13    & 1.049E+01 & 1.024      \\
  0.004 & 5.741E-06 & 1.037  &   0.14    & 1.233E+01 & 1.023      \\   
  0.005 & 2.685E-05 & 1.037  &   0.15    & 1.429E+01 & 1.023      \\ 
  0.006 & 8.666E-05 & 1.037  &   0.16    & 1.636E+01 & 1.023      \\   
  0.007 & 2.202E-04 & 1.036  &   0.18    & 2.081E+01 & 1.022      \\  
  0.008 & 4.743E-04 & 1.036  &   0.20    & 2.566E+01 & 1.022      \\ 
  0.009 & 9.058E-04 & 1.036  &   0.25    & 3.927E+01 & 1.022      \\
  0.010 & 1.579E-03 & 1.035  &   0.30    & 5.467E+01 & 1.021      \\  
  0.011 & 2.566E-03 & 1.035  &   0.35    & 7.155E+01 & 1.021      \\  
  0.012 & 3.941E-03 & 1.035  &   0.40    & 8.968E+01 & 1.021      \\ 
  0.013 & 5.780E-03 & 1.035  &   0.45    & 1.089E+02 & 1.021      \\ 
  0.014 & 8.161E-03 & 1.034  &   0.50    & 1.290E+02 & 1.021      \\
  0.015 & 1.116E-02 & 1.034  &   0.60    & 1.716E+02 & 1.022      \\  
  0.016 & 1.486E-02 & 1.034  &   0.70    & 2.168E+02 & 1.022      \\ 
  0.018 & 2.463E-02 & 1.033  &   0.80    & 2.642E+02 & 1.022      \\ 
  0.020 & 3.805E-02 & 1.033  &   0.90    & 3.133E+02 & 1.022      \\ 
  0.025 & 9.078E-02 & 1.032  &   1.00    & 3.640E+02 & 1.022      \\ 
   0.03 & 1.760E-01 & 1.031  &   1.25    & 4.961E+02 & 1.023      \\
   0.04 & 4.612E-01 & 1.030  &   1.50    & 6.344E+02 & 1.023      \\  
   0.05 & 9.154E-01 & 1.028  &   1.75    & 7.773E+02 & 1.023      \\ 
   0.06 & 1.545E+00 & 1.027  &   2.0     & 9.236E+02 & 1.024      \\  
   0.07 & 2.350E+00 & 1.027  &   2.5     & 1.221E+03 & 1.024      \\ 
   0.08 & 3.325E+00 & 1.026  &   3.0     & 1.516E+03 & 1.024      \\ 
   0.09 & 4.462E+00 & 1.025  &   3.5     & 1.796E+03 & 1.024      \\
   0.10 & 5.755E+00 & 1.025  &   4.0     & 2.050E+03 & 1.025      \\
\enddata
\tablenotetext{a}{Based on two-parameter fit, see Equation~(\ref{eq:theory}). In units of cm$^3$~mol$^{-1}$~s$^{-1}$, corresponding to the 50th percentiles of the rate probability density function. The rate factor uncertainty, $f.u.$, corresponds to a coverage probability of 68\% and is obtained from the 16th and 84th percentiles. Results are computed using 75,000 samples.}
\end{deluxetable*}

\section{Concluding summary} 
\label{sec:summary}
We reported on a comprehensive evaluation of the D(p,$\gamma$)$^3$He reaction rate, which impacts the synthesis of deuterium in the early universe. Recent studies \citep{Coc15,Pitrou2018,IC20} hinted at a tension between the observed and predicted primordial deuterium abundance values. If confirmed, this could have important implications pointing to physics beyond the standard model.

Our study took into account the largest body of data (eleven different experiments), spanning the time period of 1955--2021, including very recently published data \citep{Tisma:2019ug,Mossa20,Turkat21}. All these data, including the reported statistical and systematic uncertainties, were carefully evaluated (Appendix~\ref{sec:app}). The S-factor fitting was performed using a Bayesian model, first applied in \citet{iliadis16}, which has several advantages compared to conventional $\chi^2$ techniques. First, the assumption of Gaussian probability densities, implicit in the conventional methods, can be relaxed and forms of probability densities that best suit the circumstances can be adopted. Second, the Bayesian model incorporates all relevant uncertainties in a rigorous manner, including additional sources of statistical scatter (``extrinsic uncertainties''), which were either unknown to, or unreported by, the experimenter.

Fits were performed for two different fitting functions. One was based on microscopic nuclear theory and involved two physical model parameters. The second one was based on a cubic polynomial and was chosen to directly compare our results to those of previous studies that also employed polynomial fitting functions. We found that our reaction rates derived using the two-parameter and four-parameter fitting functions were in excellent agreement. In the temperature range most relevant for BBN ($T$ $=$ $0.1$ $-$ $1.0$~GK), our D(p,$\gamma$)$^3$He rate uncertainties are between 2.1\% and 2.5\%.

We compared our results to recently published reaction rates \citep{Mossa20,Pis21}. Compared to \citet{Mossa20}, our rates have larger uncertainties at $0.1$~GK and smaller ones at $1$~GK. As for the reason of these differences, we explored the possibility that \citet{Mossa20} have disregarded the off-diagonal matrix elements in their covariance matrix. Compared to \citet{Pis21}, our reaction rates are in better agreement, although our rate uncertainties based on a cubic polynomial S-factor exhibit a different temperature dependence.

Based on the new D(p,$\gamma$)$^3$He reaction rates from the present study, we confirm the 1.8$\sigma$ tension between the predicted and measured primordial D/H abundance ratio that was reported by \cite{Pit21,PitNat21}. {\bf At this time, the most important remaining source of uncertainty in the predicted primordial D/H value is not the current D(p,$\gamma$)$^3$He rate uncertainty (Tab.~\ref{tab:rate}), but the rate uncertainties of the D(d,n)$^3$He and D(d,p)$^3$H reactions. New measurements for those reactions in the energy range important for BBN would be highly desirable.}

\acknowledgments
We would like to thank Isabela Ti{\v s}ma for providing the numerical values listed in Table~\ref{tab:tisma}, Ofelia Pisanti for providing her reaction rates in tabular format, and Amy Nicholson for helpful comments on microscopic nuclear models. We would also like to express our appreciation to Robert Janssens for a thorough reading of the manuscript. This work was supported in part by NASA under the Astrophysics Theory Program grant 14-ATP14-0007, by the DOE, Office of Science, Office of Nuclear Physics, under grants DE-FG02-97ER41041 (UNC) and DE-FG02-97ER41033 (TUNL), and by the National Science Foundation Graduate Research Fellowship Program under Grant No.DGE-1650116. 

\appendix
\twocolumngrid

\section{Input data}
\label{sec:app}

\subsection{General aspects}
\label{sec:genaspect}
The data sets analyzed in the most recent BBN studies are summarized in Table~\ref{tab:references}. Our main selection criterion was to only consider data below a center-of-mass energy of $2.0$~MeV. Since deuterium synthesis occurs at BBN energies between $20$~keV and $300$~keV, the data at energies in excess of 2.0~MeV are irrelevant for the considerations of the present work. Experiments that reported data below 2.0~MeV are listed in the first column of Table~\ref{tab:references}. In total, we took into account 10 independent measurements, more than any other BBN study. We disregarded the data from three experiments \citep{Wol67,Gel67,Bys08}, and we will explain our reasons below.

Regarding the data that we adopted in our analysis, six experiments provided information on both statistical and systematic uncertainties \citep{War63,Sch97,Ma97,Cas02,Tisma:2019ug,Mossa20}. To this body of data we add the older results of \citet{Gri55,Gri62,Gri63,Bai70}. Since these data sets either provide no uncertainties at all or only total uncertainties, we implemented them differently into our Bayesian model compared to the other sets, as explained in the main text. Comments on individual experiments will be provided below.

\subsection{Data of \citet{Gri55}}
\label{sec:ref_gri55}
The absolute cross section normalization at a bombarding energy of 1.0~MeV is given as 50\%. In addition to this large value, the cross sections were calculated using outdated stopping powers from the early 1950's. The relative yield data, without error bars, are displayed in their Fig.~10. The text states that {\it ``no errors have been shown since the statistical errors are less than other experimental errors.''} No other information can be extracted regarding their statistical or systematic uncertainties. Cross section data are not shown in their paper, but are presented as open circles (without uncertainties) in Fig. 6 of \citet{Gri62}. Note that those open circles have been normalized to the cross section of \citet{Gri62}. We scanned the figure, extracted energies and cross sections, and converted them to center-of-mass energies and S-factors. Numerical values are presented in Table~\ref{tab:gri55}. We implemented these results into our Bayesian model by treating them as relative S-factors only (see main text).
\begin{deluxetable}{ccc|ccc}
\tablecaption{Data of \citet{Gri55}.\tablenotemark{a}
\label{tab:gri55}} 
\tablewidth{\columnwidth}
\tabletypesize{\footnotesize}
\tablecolumns{6}
\tablehead{
 $E_{cm}$   & $\sigma$  & $S$ &  $E_{cm}$   & $\sigma$  & $S$  \\
(MeV)  & ($\mu$b) & (eVb)  &  (MeV)  & ($\mu$b) & (eVb)
}
\startdata
    0.11 &     0.60 &     0.76 &   0.97 &     4.42 &     9.76\\
    0.29 &     1.91 &     2.50 &   1.00 &     4.71 &    10.59\\
    0.47 &     2.93 &     4.50 &   1.03 &     4.58 &    10.49\\
    0.65 &     3.46 &     6.13 &   1.09 &     4.67 &    11.10\\
    0.82 &     4.12 &     8.29 &   1.15 &     4.96 &    12.17\\
    0.86 &     4.00 &     8.24 &   1.19 &     5.19 &    12.98\\
    0.91 &     4.68 &     9.96 &   1.23 &     5.09 &    13.00\\
\enddata
\tablenotetext{a}{Laboratory energies and cross sections are displayed as open circles in Fig.~6 of \citet{Gri62}. The figure was scanned in the present work and the center-of-mass energies and S-factors listed here were derived.}
\end{deluxetable}

\subsection{Data of \citet{Gri62}}
\label{sec:ref_gri62}
The absolute cross sections were obtained in \citet{Gri62} by using outdated stopping powers from the 1950's. Only combined cross section uncertainties are presented in their Table I and Figure 6 (solid circles). From the published information, it is not possible to extract statistical or systematic uncertainties separately. The cross section data were converted to center-of-mass energies and S-factors by \citet{Coc15} and the results are listed in their Table~VI. We implemented these results into our Bayesian model by treating them as relative S-factors only (see main text). 

\subsection{Data of \citet{Gri63}}
\label{sec:ref_gri63}
\citet{Gri63} obtained absolute cross sections by using outdated stopping powers from the 1950's. Yields, cross sections, and S-factors are presented in their Table~I and Figures~5 and 6, respectively. The displayed error bars represent combined uncertainties, and it is not possible to extract statistical or systematic uncertainties separately. They state in the text that {\it ``The statistical errors alone are of the order of the deviations from the mean line,''} but present no numerical values. Their S-factors are listed in Table~VI of \citet{Coc15}. We implemented these results into our Bayesian model by treating them as relative S-factors only (see main text).

\subsection{Data of \citet{War63}}
\label{sec:ref_war63}
\citet{War63} measured the inverse reaction, $^3$He($\gamma$,p)D, at three different energies. The $\gamma$-ray energies and cross sections listed in their Table~II are reproduced in columns 1 and 2 of Table~\ref{tab:warren}. Using the reciprocity theorem, we calculated the energies in the $p$ $+$ $d$ center of mass and the S-factors for the forward reaction, D(p,$\gamma$)$^3$He. The values obtained are given in columns 3, 4, and 5 of Table~\ref{tab:warren}. In their Table~II, \citet{War63} also provide statistical uncertainties (which they call ``experimental errors'') and total uncertainties (which they call ``total probable errors''). We have calculated the systematic uncertainties for their four data points from the squared difference of these values, yielding 11\%, 3.6\%, 15\%, and 6.3\%. In our Bayesian model, we will adopt an average value of 10\% for the systematic uncertainty.
\begin{deluxetable}{ccccccc}
\tablecaption{Data of \citet{War63}.\tablenotemark{a}
\label{tab:warren}} 
\tablewidth{\columnwidth}
\tabletypesize{\footnotesize}
\tablecolumns{7}
\tablehead{
$E_\gamma$ & $\sigma_{\gamma,p}$  & $E_{cm}$   & $\sigma_{p,\gamma}$  & $S$  & $\Delta S_{stat}$ & $\Delta S_{tot}$ \\
(MeV)  & (mb) & (MeV) & ($\mu$b) & (eVb)  & (\%) & (\%) 
}
\startdata
  6.140 &  0.109  &   0.646  &   3.388  &  6.004   &   9.0  &  14.0 \\
  6.140 &  0.102  &   0.646  &   3.170  &  5.618   &   6.0  &  7.0 \\
  6.970 &  0.298  &   1.476  &   5.226  &  15.038  &   5.0  &  16.0 \\
  7.080 &  0.307  &   1.586  &   5.170  &  15.613  &   5.0  &  8.0 \\
\enddata
\tablenotetext{a}{Columns 1, 2, 6, and 7 are adopted from Table~II in \citet{War63}. Values in columns 3, 4, and 5 have been calculated in the present work.}
\end{deluxetable}

\subsection{Data of \citet{Bai70}}
\label{sec:ref_bai70}
\citet{Bai70} display cross sections in their Figure~1. They provide little information regarding the details of their analysis. Certainly, their absolute cross section scale rests on outdated stopping powers from the pre-1970's. We scanned their figure, extracted energies and cross sections, and converted them to center-of-mass energies and S-factors. Numerical values are presented in Table~\ref{tab:bai70}. We implemented these results into our Bayesian model by treating them as relative S-factors only (see main text).
\begin{deluxetable}{ccc|ccc}
\tablecaption{Data of \citet{Bai70}.\tablenotemark{a}
\label{tab:bai70}} 
\tablewidth{\columnwidth}
\tabletypesize{\footnotesize}
\tablecolumns{6}
\tablehead{
 $E_{cm}$   & $\sigma$  & $S$ &  $E_{cm}$   & $\sigma$  & $S$  \\
(MeV)  & ($\mu$b) & (eVb)  &  (MeV)  & ($\mu$b) & (eVb)
}
\startdata
   0.035 &     0.21 &     0.57 &  0.405 &     2.26 &     3.28\\
   0.056 &     0.42 &     0.71 &  0.426 &     2.34 &     3.45\\
   0.085 &     0.65 &     0.89 &  0.458 &     2.48 &     3.76\\
   0.254 &     1.55 &     1.96 &  0.524 &     2.70 &     4.34\\
   0.306 &     1.76 &     2.34 &  0.657 &     3.32 &     5.94\\
   0.336 &     2.00 &     2.72 &  0.724 &     3.45 &     6.48\\
\enddata
\tablenotetext{a}{Laboratory energies and cross sections are displayed as open circles in Figure~1 of \citet{Bai70}. The figure was scanned in the present work and the center-of-mass energies and S-factors listed here were derived.}
\end{deluxetable}

\subsection{Data of \citet{Ma97}}
\label{sec:ref_ma97}
The $S$-factor was scanned from Figure~9 (triangles) in \citet{Ma97} and the values were reported in Table IV of \citet{Coc15}. \citet{Ma97} state that {\it ``The systematic uncertainty for the cross sections is estimated to be $\pm$9\%, and includes the estimated errors in stopping power, detector efficiency, $\gamma$-ray angular distribution assumptions, and beam current integration.''}

\subsection{Data of \citet{Sch97}}
\label{sec:ref_sch97}
We calculated the cross section by multiplying the $A_0$ values listed in Table~II of  \citet{Sch97} by $4\pi$. The uncertainties reported in their Table~I are of statistical nature only. The authors state that their systematic uncertainty is 9\%. Our derived center-of-mass energies and S-factors are listed in Table~\ref{tab:sch}.
\begin{deluxetable}{cccc}
\tablecaption{Data of \citet{Sch97}.\tablenotemark{a}
\label{tab:sch}} 
\tablewidth{\columnwidth}
\tabletypesize{\footnotesize}
\tablecolumns{4}
\tablehead{
$E_{cm}$ (MeV) & $\sigma$ (nb) & $S$ (eVb)  &  $\Delta S_{stat}$ (eVb)
}
\startdata
  0.0100    &  7.30$\pm$0.38    &  0.2425   &  0.0125 \\
  0.0167    &  30.79$\pm$0.84   &  0.2740   &  0.0075 \\
  0.0233    &  73.26$\pm$1.38   &  0.3452   &  0.0065 \\
  0.0300    & 122.8$\pm$1.9     &  0.3974   &  0.0061 \\
  0.0367    &  176.0$\pm$2.3    &  0.4452   &  0.0057 \\
  0.0433    &  222.4$\pm$3.4    &  0.4738   &  0.0072 \\
  0.0500    & 252.6$\pm$3.4     &  0.4744   &  0.0064 \\
\enddata
\tablenotetext{a}{S-factors and uncertainties are derived from the information given in Tables~I and II of \citet{Sch97}.}
\end{deluxetable}

\subsection{Data of \citet{Cas02}}
\label{sec:ref_cas02}
The low-energy LUNA data by \cite{Cas02} are presented in their Table~I, where only statistical uncertainties are listed. The quoted systematic uncertainties range from 3.6\% at the highest measured energy to 5.3\% at the lowest one. We adopt 4.5\% for the average systematic uncertainty.

\subsection{Data of \citet{Tisma:2019ug}}
\label{sec:tisma}
The measurement of \citet{Tisma:2019ug} took place at the Joi{\v{z}}ef Stefan Institute of Ljubljana after the publication of \citet{Coc15,iliadis16}. \citet{Tisma:2019ug} measured the D(p,$\gamma$)$^3$He differential cross section at two angles and four energies from $E_{cm}$ $=$ $97$ to $210$~keV, well within the BBN energy range. The angular distribution was consistent with theory \citep{Mar16}, allowing for the extraction of total 
cross sections and $S$-factors. The results listed in Table~\ref{tab:tisma} were provided to us directly by the authors. For the systematic uncertainties, they estimated a value of 10\%. 
\begin{deluxetable}{ccc|ccc}
\tablecaption{Data of \citet{Tisma:2019ug}.\tablenotemark{a}
\label{tab:tisma}} 
\tablewidth{\columnwidth}
\tabletypesize{\footnotesize}
\tablecolumns{2}
\tablehead{
$E_{cm}$   &  $S$   &   $\Delta S_{stat}$ & $E_{cm}$   &  $S$   &   $\Delta S_{stat}$\\
(MeV)  & (eVb)  & (eVb) & (MeV)  & (eVb)  & (eVb) }
\startdata
0.097  &   0.78   &  0.13 & 0.170  &   1.57   &  0.26 \\
0.119  &   1.01   &  0.17 & 0.210  &   1.85   &  0.27 \\
\enddata
\tablenotetext{a}{Numerical values of the data shown in their Figure~10 were directly provided by the authors. The systematic uncertainty is 10\%.}
\end{deluxetable}

\subsection{Data of \citet{Mossa20}}
\label{sec:luna}
The measurement of \citet{Mossa20} took place at the LUNA-400~kV accelerator after the publication of  \citet{Coc15,iliadis16}. The experimental setup consisted of a windowless deuterium gas target and two HPGe detectors. Measured energies, S-factors, and uncertainties are reported in their Extended Data Table 1. Their reported systematic uncertainties are less than 2.7\% at all energies.

\subsection{Data of \citet{Turkat21}}
\label{sec:Dresden}
Recently, a measurement \citep{Turkat21} was performed at the 3~MV Tandetron accelerator of the Helmholtz-Zentrum Dresden-Rossendorf. The proton beam energies ranged from $400$ to $1650$~keV (or $265$ to $1094$~keV in the center of mass). Unlike the experiment of \citet{Mossa20}, they used solid targets (titanium deuteride). Gamma-rays were detected by two HPGe detectors placed at 55$^{\circ}$ and 90$^{\circ}$. Their $S$-factors can be found in Table~II of \citet{Turkat21}, together with statistical and systematic uncertainties. Their absolute normalization is subject to a reported global uncertainty of 12\%, with additional scaling uncertainties of 5\% or 8\% depending on the target (see Table~III in \citet{Turkat21}). We implemented their data into our Bayesian model assuming an overall systematic uncertainty of $14$\%.

\subsection{Disregarded data}
\label{sec:ref_not}
\cite{Wol67} measured the D(p,$\gamma$)$^3$He cross section at laboratory proton energies between 2~MeV and 12~MeV. Only three of their data points pertain to center-of-mass energies below 2~MeV. We decided to disregard these data points for the following reasons. The cross sections are shown in their Figure~5, but no numerical values are reported. From their figure it is apparent that the scatter of the data points is much less than the size of the displayed error bars, implying that systematic errors dominate the total uncertainties. It is not possible to derive statistical and systematic uncertainties separately, based on the information presented in their publication.

\citet{Gel67} reported only differential cross sections measured at 90$^\circ$. Results are displayed in their Figure~2, but no numerical values are presented in tables. Seven of their data points are located below a center-of-mass energy of 2~MeV. They state that the absolute cross section scale {\it ``is obtained by normalizing the gamma-ray yield''} to a theoretical prediction (i.e., Gunn-Irving model). Little information is given about the meaning of the displayed error bars, and it is not possible to extract statistical and systematic uncertainties reliably. While we could have implemented these data points into our Bayesian model by treating them as relative S-factors, we decided to disregard these results. Taking them into account would have required us to introduce additional assumptions and uncertainties for angular distribution corrections. 

In previous analyses \citep{Coc15,iliadis16}, the low-energy data from Table~3 of \citet{Bys08} were used because both statistical and systematic uncertainties ($<$8\%) were reported. However, the statistical uncertainties are rather large ($\approx$30\%). Furthermore, electron screening could be important at these very low energies, depending on the target used. The same group followed up on their 2008 experiment with measurements over a wider energy range, with higher statistics, and different types of targets \citep{Bys14a,Bys14b,Bys15}. Unfortunately, these latter works provide no information on statistical or systematic uncertainties. Furthermore, they found that the cross section is affected by electron screening when titanium deuteride is used as a target, but no effect was observed with zirconium deuteride or a frozen D$_2$O target. Because this issue remains unresolved at this time, we disregarded these data.

\citet{Zylstra} reported on an experiment performed at the OMEGA laser facility, which produced an inertially confined plasma of a H and D mixture. They measured $\gamma$ rays from the \dpg\ reaction and neutrons from the \ddn\ reaction, and determined from the reaction yields the ratio of reactivities (i.e., the Maxwellian-averaged cross sections). Their extracted $S$-factor is $S$ $=$ $0.429\pm0.026$(stat) $\pm$ $0.072$(sys)~eVb at $E_{cm}$ $=$ $16.35\pm0.40$~keV. Even though this technique is promising, it does not represent a direct measurement of the $S$-factor or cross section. The $S$-factor value had to be unfolded from the measured reactivity, assuming that the plasma temperature can be accurately determined from the neutron spectrum. In addition, the extracted value does not represent an independent absolute $S$-factor measurement, because it was determined relative to the \ddn\ $S$-factor of \citet{Bos92}. Therefore, we do not include this single data point in the analysis.

\section{Comments on the Mossa {\it et al.} and Pisanti {\it et al.} S-factors}
\label{sec:app2}
\begin{figure}
\includegraphics[width=1\linewidth]{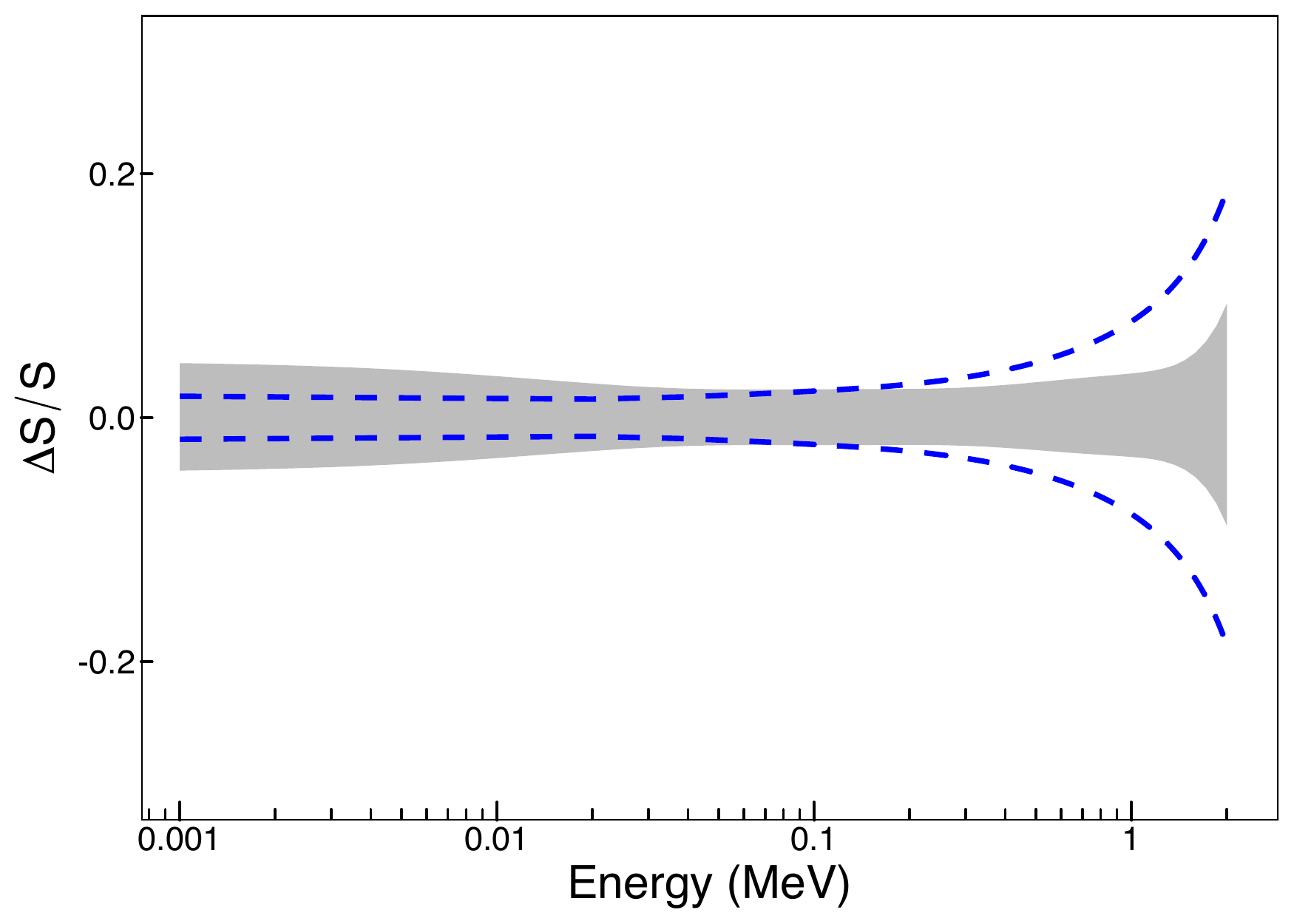}
\caption{Relative S-factor uncertainty obtained using a cubic polynomial fit function. (Gray band) Result of present four-parameter fit function, i.e., the same result as shown in the lower panel of Figure~\ref{fig:sfactor2}. (Blue dashed lines) Result of \citet{Mossa20}, according to Equation~(\ref{eq:mos}), which most likely disregards the off-diagonal elements of the covariance matrix (see text).
}
\label{fig:covar}
\end{figure}

\citet{Mossa20} provide their $S$-factor, including uncertainties, in their Equations~(2) and (3), 
\begin{align}\label{eq:mos2}
S(E) = & ( 0.2121 + 5.973\times10^{-3}\,E \notag \\ 
& + 5.449\times10^{-6}\,E^2  \notag \\
&  - 1.656\times10^{-9}\,E^3 ) ~ \text{eV\,b},
\end{align}
\begin{align}\label{eq:mos}
(\Delta S(E))^2 = & (1.4 \times 10^{-5} + 2.97 \times 10^{-8}\,E^2 \notag \\
& + 4.80 \times 10^{-13}\,E^4 \notag \\
& + 1.12 + \times 10^{-19}\,E^6) ~ \text{eV$^2$\,b$^2$},
\end{align}
with the center-of-mass energy given in units of kilo electron volts. Their fractional S-factor uncertainty, $\Delta S/S$, is depicted by the blue dashed lines in Figure~\ref{fig:covar}. For comparison, our results using the four-parameter fit function are displayed as the gray band (68\% coverage probability), which is the same as the one shown in Figure~\ref{fig:sfactor2}. The reaction rate in the Extended Data Table~2 of \citet{Mossa20} can be reproduced by integrating these equations and is displayed in Figure~2 of \citet{Pit21}. 

The data sets taken into account in the fits of \citet{Mossa20} and \citet{Pis21} are almost identical, except that the latter work considered in addition the data of \citet{Tisma:2019ug}. Specifically, \citet{Pis21} state that their inclusion of the data of \citet{Tisma:2019ug} (see Table~\ref{tab:references}) yields only small differences in the S-factor compared to the results given in \citet{Mossa20}. 

The mean S-factor of \citet{Pis21} is given by
\begin{align}
S(E) & = (0.2121 + 5.975\,E + 5.463\,E^2 \notag \\
& - 1.665\,E^3) \times 10^{-6} ~ \text{MeV\,b}, \label{eq:pis}
\end{align}
with the energy in mega electron volts, and is indeed very similar to the mean S-factor of \citet{Mossa20} (see Equation~(\ref{eq:mos2})). \citet{Pis21} did not report their S-factor uncertainties directly, but provided their covariance matrix, given by
\begin{align}
& \text{cov}(S_i,S_j) = \notag \\
& 10^{-15} \times \left(
\begin{array}{cccc}
0.0140   & -0.378  & 1.07   & -0.462  \\
-0.378   & 29.5    & -90.0  & 39.5  \\
1.07     & -90.0   & 479.0  & -230.0 \\
-0.462   & 39.5    & -230.0 & 112.0 
\label{eq:covar}
\end{array}
\right).
\end{align}
Notice that the diagonal elements are very close in value to the coefficients of Equation~(\ref{eq:mos}). Considering the near numerical equality in the mean S-factors of \citet{Mossa20} and \citet{Pis21} (Equations~(\ref{eq:mos2}) and (\ref{eq:pis}), respectively), this indicates that the S-factor presented in \citet{Mossa20} most likely did not take into account the off-diagonal terms in the covariance matrix (i.e., they seem to have disregarded the correlations between the polynomial coefficients, $S_i$). Also, notice that it is not clear how to extract S-factor uncertainties from the information reported in \citet{Pis21}.

\bibliographystyle{aasjournal}
\bibliography{ref}
\end{document}